# Minimum Distance Estimation for Robust High-Dimensional Regression

Aurélie C. Lozano[1] and Nicolai Meinshausen[2]

[1]IBM T.J. Watson Research Center, USA
[2]Department of Statistics, University of Oxford, UK

July 11, 2013

### Abstract

We propose a minimum distance estimation method for robust regression in sparse high-dimensional settings. The traditional likelihood-based estimators lack resilience against outliers, a critical issue when dealing with high-dimensional noisy data. Our method, Minimum Distance Lasso (MD-Lasso), combines minimum distance functionals, customarily used in nonparametric estimation for their robustness, with $\ell_1$-regularization for high-dimensional regression. The geometry of MD-Lasso is key to its consistency and robustness. The estimator is governed by a scaling parameter that caps the influence of outliers: the loss per observation is locally convex and close to quadratic for small squared residuals, and flattens for squared residuals larger than the scaling parameter. As the parameter approaches infinity, the estimator becomes equivalent to least-squares Lasso. MD-Lasso enjoys fast convergence rates under mild conditions on the model error distribution, which hold for any of the solutions in a convexity region around the true parameter and in certain cases for every solution. Remarkably, a first-order optimization method is able to produce iterates very close to the consistent solutions, with geometric convergence and regardless of the initialization. A connection is established with re-weighted least-squares that intuitively explains MD-Lasso robustness. The merits of our method are demonstrated through simulation and eQTL data analysis.

Keywords: Robust estimation, high-dimensional variable selection, sparse learning.

## 1 Introduction

We address the problem of robust sparse estimation in high-dimensional regression. Sparse linear models allow for simultaneous model estimation and variable selection. They have become very popular tools to analyze the high-dimensional data that is prevalent in many domains such as genomics [40], neuroimaging [18, 37], and economics [13]. A widely used approach to sparse learning is via sparsity-inducing regularization. A well known example is the Lasso [33], which employs $\ell_1$-penalized least-squares to identify a parsimonious subset of predictors. Beyond Lasso, various structured penalties have been proposed that reflect the underlying structural information among the predictors. For instance the Group Lasso [41] enforces group sparsity via the $\ell_1/\ell_q$ norm $(q > 1)$, the Path Coding Penalties [24] and Graph Lasso [17] deal with applications where the variables reside in a graph, the Fused Lasso [34] enforces sparsity in both the coefficients and their successive differences for settings where variables are ordered in some meaningful way and a locally constant coefficient profile is desirable. Much attention has been devoted recently to the study of these structured norms and their theoretical properties [25, 26], and to devising efficient algorithms for large scale problems [5].

The issue of robustness, however, has been largely overlooked in the sparse learning literature, while this aspect is critical when dealing with high dimensional noisy data. Traditional likelihood-based estimators (including Lasso and Group Lasso) are known to lack resilience to outliers and model misspecification. Despite this fact, there has been limited focus on robust sparse learning methods in high-dimensional modeling. Relevant penalized regression methods include the "extended" Lasso formulation [28] which employs the traditional squared error but incorporates an additional sparse error vector into the model so as to account for corrupted observations, and the LAD-Lasso [38], which uses the least absolute deviation combined with an $\ell_1$ penalty. Note that the least absolute deviations estimate also arises as a maximum likelihood estimate if the errors have a Laplace distribution, hence the aforementioned approaches can still be viewed as likelihood-based.

Departing from likelihood-based methods, we propose a penalized minimum distance criterion for robust and consistent estimation of sparse high dimensional regression. Our approach is motivated by minimum distance estimators [39], which are popular in nonparametric methods and have been shown to exhibit excellent robustness and efficiency properties [9, 12]. Their use for parametric estimation has been discussed in [7, 31]. More specifically, we propose the Minimum Distance Lasso (MD-Lasso) estimator, which is derived from the integrated squared error distance between the model and the "true" distribution, and imposes sparse model structure via $\ell_1$ penalty. The MD-Lasso loss, taken as a function of a single observation, acts similarly to the squared-loss if the residual squared-error of that observation is small, while the loss becomes flat as the squared-error becomes large. This ensures that the contributions of large outliers to the overall loss are capped. Overall the MD-Lasso loss is invex[1] and locally convex. The extent of the local convexity region and the capping of outliers are both governed by a scaling parameter of our estimator, against which the residual squared error of each observation is being compared. In the extreme case where the scaling parameter goes to zero, only the most "trusted" observation is taken into account, while as the scaling parameter goes to infinity, the estimator becomes equivalent to the traditional Lasso estimator with the same amount of regularization on the $\ell_1$ penalty. Our analysis shows that the tradeoff between convexity and robustness, as controlled by the scaling parameter, is, understandably, essential in securing both robustness and consistency of the estimator.

Our results demonstrate that the MD-Lasso enjoys fast convergence rates in high dimensional settings under mild conditions on the model error distribution in relation to the scaling parameter. These conditions are much less restrictive than the traditional sub-gaussian assumption, and cover a broad class of heavy-tailed distributions. They also have an intuitive interpretation which is similar to the classical notion of breakpoint point [16], namely the MD-Lasso is able to tolerate a certain percentage of outliers (with arbitrarily large residuals) and still retain the fast convergence rates. Our consistency results hold for any of the solutions in the local convexity region around the true parameter, and in many cases for *any* solution globally. Remarkably, even in the event where local minima might exist outside of the convex region, we show that a simple composite gradient algorithm yields an estimate that is very close to any consistent solution, with geometric numerical convergence, and regardless of the initialization.

---

[1] A function $f$ is invex if it is differentiable and there exists a vector-valued function $g$ such that $|f(\bm{v}) - f(\bm{u})| \leq \langle \nabla f(\bm{u}), g(\bm{v}, \bm{u}) \rangle$, for all $\bm{u}, \bm{v}$. A function is invex if and only if every stationary point is a global minimum [8].

Our work is in line with the recent manuscript of [22] on non-convex M-estimators. Both works leverage the statistical and optimization results obained in [26, 1] for convex M-estimators, noting that global convexity is actually not required for their analysis to be applicable. The emphasis, however, is different: [22] focuses on noisy/missing *covariates* and non-convex penalties (they consider settings with sub-gaussian errors only, and loss functions derived from Maximum-Likelihood formulations), while our work targets outliers/noise in the *error* term. Hence both works complement each other nicely. An important part of our study is concerned with showing that gradient boundedness and local convexity (or RSC-type conditions) hold with high probability under heavy-tailed error distributions for the MD-Lasso loss.

We shed further light into the robustness of MD-Lasso by establishing its connection with a form of iteratively weighted $\ell_1$-penalized least-squares regression (namely the traditional Lasso) where the weights assigned to the observations can be interpreted in terms of their likelihood.

The performance of our estimator is demonstrated on simulation data under various error distributions, in comparison to the traditional Lasso, LAD-Lasso, and Extended Lasso. This study also confirms that outliers and/or heavy-tailed noise can severely influence the variable selection accuracy of existing sparse learning methods. Experiments on real eQTL data further illustrate the usefulness of our approach.

The manuscript is organized as follows. The problem formulation and notation are introduced in Section 2. Section 3 is devoted to the MD-Lasso estimator, its derivation from a minimum distance criterion, and its geometry. The statistical consistency results and convergence rates are shown in Section 4. The composite gradient method for efficient and scalable optimization is presented in Section 5, along with theoretical guarantees on the quality of its iterates. Empirical results are described in Sections 6 and 6.2. All the technical proofs are collected in the Appendix.

## 2 Problem Formulation and Notation

Let $\boldsymbol{X} \in \mathbb{R}^{n \times p}$ denote the predictor matrix, whose rows are $p$-dimensional variable vectors observed for $n$ training examples. Denote by $\boldsymbol{X}_i \in \mathbb{R}^p$ the vector formed by $i^{\text{th}}$ observation across all variables. Denote by $\boldsymbol{X}^j \in \mathbb{R}^n$ the vector formed by the observations for the $j^{\text{th}}$ variable. Denote by $X_i^j \in \mathbb{R}$ the entry in matrix $\boldsymbol{X}$ corresponding to the $i^{\text{th}}$ observation for the $j^{\text{th}}$ variable. Similarly let $\boldsymbol{Y} \in \mathbb{R}^n$ denote the response vector, and $Y_i$ it's $i$-th observation.

Consider the general regression model:

$$\boldsymbol{Y} = \boldsymbol{X}\boldsymbol{\beta}^\star + \boldsymbol{\eta}, \tag{1}$$

where $\boldsymbol{\beta}^\star \in \mathbb{R}^p$ is the coefficient vector one wishes to estimate, and $\boldsymbol{\eta} \in \mathbb{R}^n$ is the error term, and for simplicity we assume that the data have been standardized so that we need not consider intercept terms.

We address the sparse estimation of coefficient vector $\boldsymbol{\beta}^\star$ via $\ell_1$-penalized loss minimization.

Specifically, we consider estimators of the form

$$\hat{\boldsymbol{\beta}}_{\lambda_n} = \arg\min_{\boldsymbol{\beta}} \; L(\boldsymbol{\beta}) + \lambda_n \|\boldsymbol{\beta}\|_1, \qquad (2)$$

where the loss function $L$ measures the goodness-of-fit on the response and $\lambda_n$ is the regularization parameter for the $\ell_1$ penalty. However our framework is readily applicable to other sparsity-inducing penalties such as the group or fused Lasso.

Using likelihood-based loss functions such as squared loss is a common practice in estimating and exploring the sparsity structure of the unknown parameters for model (1), whereby $L(\boldsymbol{\beta})$ is derived from a product of probability density functions (p.d.f.). However, the likelihood-based estimators are known to lack resilience to outliers and model misspecification. In contrast, the minimum distance estimators [39] often used in nonparametric function estimation show excellent robustness properties [9, 12]. This motivates our proposed MD-Lasso estimator.

## 3 The Minimum Distance Lasso Estimator

We begin this section by presenting how the MD-Lasso objective can be rigorously derived from a minimum distance criterion.

### 3.1 Minimum Distance Estimation

Here we treat response $Y$ and predictors $\boldsymbol{X}$ as random variables, where $Y \in \mathbb{R}$ and $\boldsymbol{X} \in \mathbb{R}^p$. We first apply the Integrated Squared Error to the conditional distribution of response $Y$ given the predictors $\boldsymbol{X}$. This leads to an $\ell_2$ distance between the true conditional distribution $f(Y|\boldsymbol{X})$ and the parametric distribution function $f(Y|\boldsymbol{X}; \beta)$ as follows

$$\begin{aligned} d(\beta) &= \int [f(Y|\boldsymbol{X};\boldsymbol{\beta}) - f(Y|\boldsymbol{X})]^2 \, dY \qquad (3) \\ &= \int f^2(Y|\boldsymbol{X};\boldsymbol{\beta}) dY - 2\int f(Y|\boldsymbol{X};\boldsymbol{\beta}) f(Y|\boldsymbol{X}) dY + \int f^2(Y|\boldsymbol{X}) dY \\ &= \int f^2(Y|\boldsymbol{X};\beta) dY - 2\mathbb{E}[f(Y|\boldsymbol{X};\boldsymbol{\beta})] + \text{constant}. \end{aligned}$$

where $\int f^2(Y|\boldsymbol{X}) dY$ is a constant independent of $\boldsymbol{\beta}$.

We remark that minimum distance estimators originally involved distances between cumulative distribution functions [39], but the notion was subsequently broadened to encompass distances between probability density functions [9, 12, 31]. We consider the latter, which is easier to work with, and is also more natural in the context of linear regression.

Note that we assume a parametric family for the model while using a nonparametric criterion (the Integrated Squared Error) to measure goodness of fit. From the perspective of the loss function, the Integrated Squared Error is a more robust measure of the goodness-of-fit compared to likelihood-based loss functions. It can match the model with the largest portion of the data because the integration in (3) accounts for the whole range of the squared loss function.

To derive our estimator, we assume that $f(Y|\boldsymbol{X};\boldsymbol{\beta})$ is the p.d.f. of multivariate normal $\mathcal{N}(\boldsymbol{X}'\boldsymbol{\beta}, \sigma^2)$. However it is important to note that our methodology and theoretical results go well beyond the normal assumption for the error. $f(Y|\boldsymbol{X};\boldsymbol{\beta}) \equiv f(Y - \boldsymbol{X}'\boldsymbol{\beta})$ because of the conditional distribution assumption, and it holds that $\int f^2(Y|\boldsymbol{X};\boldsymbol{\beta})dY = 1/(2\pi^{1/2}\sigma)$. Since $\eta_i = Y_i - \boldsymbol{X}'_i\beta, i = 1, \ldots, n$ are independently and identically distributed, one can consider estimating $\mathbb{E}[f(Y|\boldsymbol{X};\boldsymbol{\beta})]$ by the empirical mean $n^{-1}\sum_{i=1}^{n} f(Y_i|\boldsymbol{X}_i;\boldsymbol{\beta})$. Such an approximation technique has also been used for Gaussian mixture density estimation [31]. Disregarding the constant terms that are independent of $\boldsymbol{\beta}$ we can write the resulting empirical criterion as

$$d_n(\beta) = -\frac{2}{n}\sum_{i=1}^{n} f(Y_i|\boldsymbol{X}_i;\boldsymbol{\beta})$$

$$= -\frac{2}{n}\sum_{i=1}^{n} \frac{1}{\sqrt{2\pi\sigma^2}} \exp(-\frac{1}{2\sigma^2}(Y_i - \boldsymbol{X}'_i\boldsymbol{\beta})^2). \quad (4)$$

Rather than directly minimizing $d_n$ we will aim to minimize, equivalently,

$$-\log(-d_n(\boldsymbol{\beta})) = -\log\left(\frac{2}{n\sqrt{2\pi\sigma^2}}\sum_{i=1}^{n} \exp\left(-\frac{1}{2\sigma^2}(Y_i - \boldsymbol{X}'_i\boldsymbol{\beta})^2\right)\right)$$

$$= -\log\left(\sum_{i=1}^{n} \exp\left(-\frac{1}{2\sigma^2}(Y_i - \boldsymbol{X}'_i\boldsymbol{\beta})^2\right)\right) + C,$$

where $C$ is a function of $\sigma$ and $n$ but is independent of $\boldsymbol{\beta}$. As $\sigma$ is unknown we consider instead

$$L(\boldsymbol{\beta}) = -c\log\left(\sum_{i=1}^{n} \exp\left(-\frac{1}{2c}(Y_i - \boldsymbol{X}'_i\boldsymbol{\beta})^2\right)\right)$$

where $c$ is a scaling parameter. Plugging the resulting loss in (2) yields the MD-Lasso problem:

$$\hat{\boldsymbol{\beta}}_{\lambda_n} = \operatorname*{argmin}_{\beta} \left(-c\log\left[\sum_{i=1}^{n} \exp(-\frac{1}{2c}(Y_i - \boldsymbol{X}'_i\boldsymbol{\beta})^2)\right] + \lambda_n\|\boldsymbol{\beta}\|_1\right). \quad (5)$$

**Remarks.** We can already gain some intuition on the robustness of MD-Lasso by considering the ratio between data and model probability density functions: $f(Y|\boldsymbol{X})/f(Y|\boldsymbol{X};\boldsymbol{\beta})$. An outlier in the data may drive this ratio to infinity, in which case the log-likelihood becomes infinite as well. In contrast, the difference between $f(Y|\boldsymbol{X}) - f(Y|\boldsymbol{X};\boldsymbol{\beta})$ as in (3) is always bounded. This property makes the $\ell_2$-distance a favourable choice when dealing with outliers. A similar argument can be applied to the problem of density estimation and explains why the $\ell_2$-distance is also very well suited for this problem (e.g. see [32]). A more heuristical intuition comes from noting that in (5) the logarithm is applied to a *sum* of probability density functions, in contrast to the likelihood-based estimators which involve a *product*: the sum should be more robust to noise and outliers, as often encountered in high-dimensional data.

## 3.2 The geometry of the MD-Lasso estimator

The geometry of MD-Lasso is worth examining, as it provides some key insights on the estimator's robustness and the theoretical conditions required for fast convergence rates. The MD-Lasso loss, taken as a function of a the residual error for a *single* observation with the contributions from the other observations fixed, is depicted in Figure 1, for various values of the scaling parameter $c$, along with the squared loss and the absolute loss. In the figure, the MD-Loss has been translated w.r.t. the y-axis for ease of comparison. From Figure 1 we can see that the MD-Lasso loss acts similarly to the squared-loss if the residual squared-error of that observation is small, while the loss becomes flat as the squared-error becomes large. This insures that the contributions of large outliers to the overall loss are capped. The range of the similarity to the squared loss is governed by the scaling parameter $c$ of the MD-Lasso estimator, against which the residual squared error of each observation is being compared. Intuitively, the scaling parameter can thus be interpreted as a cut-off on what is an acceptable range for the error.

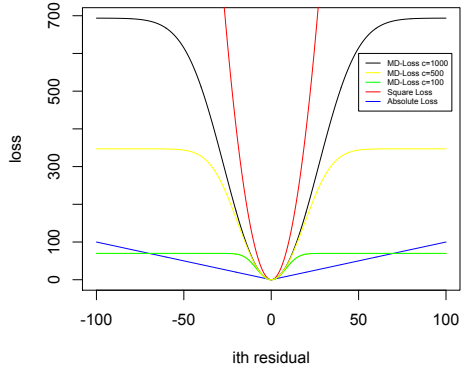

Figure 1: The MD-Lasso loss, squared loss and absolute loss, as a function of the residual error of a single observation.

The MD-Lasso loss over all observations is depicted in Figure 3.2 as a function of the regression parameter vector $\boldsymbol{\beta}$, for an illustrative examples with dimensionality $p = 2$ and just a single relevant predictor, namely $\boldsymbol{\beta}^\star = (\beta_1^\star, 0)$. As shall be formally discussed in Section 4, the MD-Lasso loss is invex and locally convex, yet it is globally non-convex. The extent of the local convexity region is controlled by the scaling parameter $c$. As the parameter increases, the convexity region becomes larger, and so does the proportion of observations whose squared-error are below the scaling parameter. The robustness, however, becomes weaker, as instances with larger error are allowed to significantly contribute to the overall loss. If the scaling parameter becomes too small, the proportion of observations with squared error below the scaling parameter becomes too small and compromises the convexity of the *overall* loss with respect to the model coefficients, a property that is needed to yield fast convergence rates.

**Limit cases.** As the parameter $c$ goes to infinity the MD-Lasso estimator becomes equivalent to the traditional Lasso estimator with the same amount of regularisation $\lambda_n$. Indeed as $x \to 0$, we have $\exp(x) \sim 1 + x$ and $\log(1 + x) \sim x$. Therefore as $c \to \infty$, the MD-Lasso estimator is identical

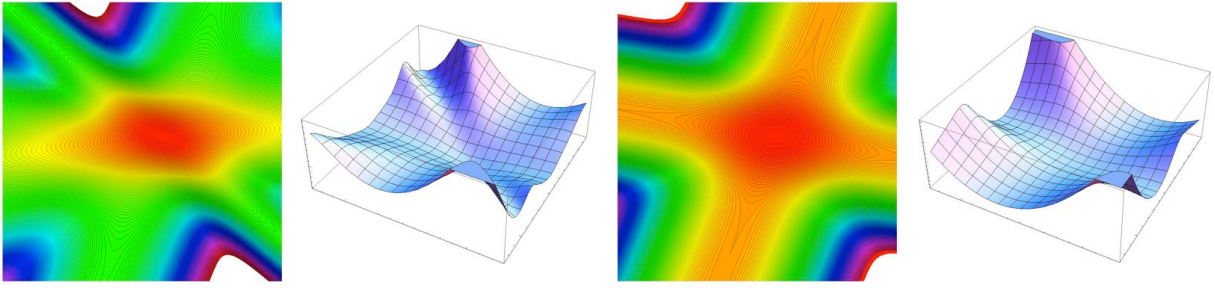

Figure 2: Contour plot and graph of the MD-Lasso loss for illustrative examples where the dimensionality $p = 2$, and sample size $n > p$ (two plots on the left), and $n = p$ (two plots on the right)

the minimizer of
$$\frac{1}{2n}\sum_{i=1}^{n}(Y_i - \boldsymbol{X}_i'\boldsymbol{\beta})^2 + \lambda_n\|\boldsymbol{\beta}\|_1.$$

On the other extreme, for $c \to 0$, the MD-Lasso is equivalent to the minimizer of
$$\min_{i=1,\ldots,n} \frac{1}{2}(Y_i - \boldsymbol{X}_i'\boldsymbol{\beta})^2 + \lambda_n\|\boldsymbol{\beta}\|_1.$$

In the latter case, only the observation with smallest residual error is taken into account, while the other observations are being discarded. This setting can thus be viewed as an extreme case of trimmed Lasso regression, where all but one observation are trimmed out.

**Non-convexity and Robustness.** To illustrate the limitations of convex loss functions and the appropriateness of non-convex loss functions with respect to robustness, it is worthwhile to recall the notion of *influence function* from robust statistics [15]. Consider the loss $L$ as a function of the residual $r_i$ of a single sample. The influence function represents the rate of change in $L$ upon a small amount of contamination on $r_i$, and thus measures the effect of the size of a residual on the loss. Specifically, for loss functions induced by log-concave densities (e.g. the squared and absolute losses are induced by the Gaussian and Laplace distributions respectively) the influence function is identical to the derivative of the loss with respect to the residual. For squared error loss, the influence function is given for example by $I(r_i) = r_i$ and for the absolute loss it is $I(r_i) = \text{sign}(r_i)$. In our case the influence function can be written as
$$I(r_i) = \frac{r_i}{1 + d\exp\left(\frac{r_i^2}{2c}\right)},$$

where $d > 0$ is a constant due to the contribution from other observations. We can see that in contrast to the case of log-concave densities, the influence function of the MD-Lasso loss is *redescending* as $r_i$ becomes large, which signifies that large residuals are basically ignored. The so-called *redescending* behavior is a desirable property for robustness, which clearly cannot be achieved by convex loss functions. We refer the reader to [15] for a review of influence-function approaches to robust statistics, including redescending influence functions.

It is also important to realize that the negative log-likelihood functions of heavy tailed distributions (e.g. Student's t and Cauchy) are non-convex. For instance for a Student's t error model with $\nu$ degrees of freedom, the loss becomes $L(r_i) = \log(1 + r_i^2/\nu)$. It thus appears that non-convexity is clearly needed to accommodate large outliers or significantly noisy data. See [2] for additional pertinent points elucidating the need for non-convex loss functions to achieve robustness.

## 4 Main Results

In this section we establish the conditions for consistency and fast convergence rates of the MD-Lasso estimator under the high-dimensional setting ($p \gg n$). The proofs of our results are all relegated to the appendix. Consistency and fast convergence rates can be secured thanks to two key properties: (i) the *restricted strong convexity* of the loss $L$ in the neighborhood of the true model parameter vector and (ii) the *gradient boundedness* at the true model parameter vector. The importance of these two properties was first identified in [26]. Before defining and establishing them, we introduce some notation and state the assumptions required by our analysis.

**Notation.** Define $\kappa_\gamma$ to be the cumulative distribution function of $|\eta_i|$ such that for all $\gamma \geq 0$,

$$\kappa_\gamma := P(|\eta_i| \geq \gamma).$$

Let $S$ denote the set of indices corresponding to the support of the true coefficient vector $\boldsymbol{\beta}^\star$. Writing $\boldsymbol{\Delta}_S$ for the projection of a vector $\boldsymbol{\Delta} \in \mathbb{R}^p$ onto indices $S$, and $\boldsymbol{\Delta}_{S^C}$ for the projection onto the complement of $S$, define the cone

$$C(S) := \{\boldsymbol{\Delta} \in \mathbb{R}^p \mid \|\boldsymbol{\Delta}_{S^C}\|_1 \leq 3\|\boldsymbol{\Delta}_S\|_1\}.$$

**Assumptions.** We make the following assumptions throughout.
[A1] Bounded predictors: there exists $M < \infty$ such that $|X_i^j| \leq M$ for all $i = 1, \ldots, n$ and $j = 1, \ldots, p$.
[A2] The error terms $(\eta_i)_{i=1}^n$ form a sequence of independent and identically distributed random variables, with zero-mean or, if the mean is undefined, a probability density function symmetric around zero.
[A3] The design matrix $\boldsymbol{X}$ satisfies the following Restricted Eigenvalue condition

$$\frac{\|\boldsymbol{X}\boldsymbol{\Delta}\|_2^2}{n} \geq \kappa_{RE}\|\boldsymbol{\Delta}\|_2^2, \text{for all } \boldsymbol{\Delta} \in C(S) \tag{6}$$

with $\kappa_{RE} > 0$. Assumption [A1] could be relaxed in some ways but we are mainly interested here in robustness with respect to outliers in the target. The second assumption [A2] is weak since it allows arbitrarily heavy tails in the error distribution, while the last assumption [A3] is standard, see for example [10].

### 4.1 Gradient Boundedness Property

The following results provide upper-bounds on the $\ell_\infty$-norm of the gradient of the MD-Lasso loss evaluated at $\boldsymbol{\beta}^\star$. These bounds are the most important part when establishing rates of convergence.

**Lemma 1** *Under Assumptions [A1] and [A2] for any $\gamma \leq \sqrt{c}$ let $\xi_{c,\gamma} \geq 0$ be given by*

$$\xi_{c,\gamma}^2 = M^2 \kappa_1^{-2} [(1 - 2\kappa_\gamma)\gamma^2 \exp(-\gamma^2/c) + 2c\kappa_\gamma/e] e^{1/c},$$

*Then, for some positive constants $\alpha_1, \alpha_2$,*

$$P\left(\|\nabla L(\boldsymbol{\beta}^\star)\|_\infty \leq \xi_{c,\gamma} \sqrt{\frac{\log p}{n}}\right) \geq 1 - \alpha_1 \exp(-\alpha_2 \xi_{c,\gamma}^2 \log p).$$

A proof is given in the appendix.

**Lemma 2** *Under Assumptions [A1] and [A2] let $\zeta_c \geq 0$ be given by*

$$\zeta_c^2 = 4M^2 \kappa_1^{-2} E[\eta_i^2 e^{-\eta_i^2/c}] e^{1/c}.$$

*Then, for some positive constants $\alpha_1', \alpha_2', \alpha_3'$,*

$$P\left(\|\nabla L(\boldsymbol{\beta}^\star)\|_\infty \leq \zeta_c \sqrt{\frac{\log p}{n}}\right) \geq 1 - \alpha_1' \exp\left(-\alpha_2' \zeta_c^2 \log p \frac{1 - \alpha_3'\sqrt{c \log(p)/n}}{1 + \alpha_3'\sqrt{c \log(p)/n}}\right).$$

A proof is given in the appendix.

**Remarks.** We note that Lemma 1 rests on establishing the bounded differences property of the gradient coordinates, based on whether or not the amplitude of the error $\eta_i$ exceeds $\sqrt{c}$. Lemma 2 employs Bernstein's inequality [20], noting that the variance of $\eta_i \exp(-\eta_i^2/(2c))$ is always well-defined regardless of whether or not the variance of $\eta_i$ exists.

As $c \to \infty$ the bound in Lemma 1 becomes vacuous: it essentially scales with $\sqrt{c}$. For heavy-tailed distributions, this is an accurate indication that large values of $c$ are not an option, as this would essentially mean giving up on the robustness property of the estimator. For lighter-tailed distributions for which the variance of $\eta_i$ exists (and is finite), Lemma 2 is preferred. Since $\xi_c \to 4M^2 \kappa_1^{-2} E[\eta_i^2]$ for $c \to \infty$ (by the monotone convergence theorem), Lemma 2 yields finite upper-bounds if $c \to \infty$ but with a rate depending on $n$ such that $c \log(p)/n \to 0$. We present below some specific examples illustrating this interesting fact. If the variance of $\eta_i$ is undefined Lemma 1 yields tighter bounds for large values of $c$.

**Examples.** GAUSSIAN ERRORS. If the error terms $\eta_i, i = 1, \ldots, n$ follow a Gaussian distribution $N(0, \sigma^2)$ and $c$ is finite, then Lemma 2 implies that with high probability

$$\|\nabla L(\boldsymbol{\beta}^\star)\|_\infty \leq 2M\sigma \frac{\left(\frac{c}{2\sigma^2+c}\right)^{3/4} e^{1/2c}}{\kappa_1} \sqrt{\frac{\log p}{n}}.$$

If $c \to \infty$ while $c \log(p)/n \to 0$, we recover the condition for the traditional Lasso (up to a constant factor) namely:

$$\|\nabla L(\boldsymbol{\beta}^\star)\|_\infty \leq 2M\sigma \sqrt{\frac{\log p}{n}}$$

This is consistent with the fact that the MD-Lasso estimator yields the traditional Lasso estimator as $c \to \infty$.

LAPLACE-DISTRIBUTED ERRORS. If the error terms $\eta_i$, $i = 1, \ldots, n$ follow a Laplace distribution with scale parameter $b$, and $c$ is finite, then Lemma 2 implies that with high probability

$$\|\nabla L(\boldsymbol{\beta}^\star)\|_\infty \leq \frac{2Me^{1/2c}}{\kappa_1}\sqrt{-\frac{c^2}{4b^2} + \frac{\sqrt{2\pi}}{b}\left(\frac{c}{2}\right)^{3/2} e^{\frac{c}{4b^2}}\left(1 + \frac{c}{2b^2}\right)\bar{F}\left(\frac{1}{b}\sqrt{\frac{c}{2}}\right)}\sqrt{\frac{\log p}{n}},$$

where $\bar{F}(\cdot)$ denotes the tail probability function of the standard normal distribution.

If $c \to \infty$ while $(c \log p)/n \to 0$, Lemma 2 together with the monotone convergence theorem yields the condition

$$\|\nabla L(\boldsymbol{\beta}^\star)\|_\infty \leq \sqrt{8}\frac{Mb}{\kappa_1}\sqrt{\frac{\log p}{n}}.$$

We will use these gradient bounds to show fast convergence rates of the MD-loss under potentially heavy-tailed distributions.

## 4.2 Restricted strong convexity

The following lemma states conditions that guarantee the strong convexity of the MD-Lasso loss in a restricted neighborhood of the true model coefficients $\boldsymbol{\beta}^\star$.

**Lemma 3** *Under Assumptions [A1] and [A3] for any $\mu < \sqrt{c}/(8M\sqrt{s})$, where $s = |S|$, consider the set $K(S, \mu) = \{\boldsymbol{\Delta} \in C(S) : \|\boldsymbol{\Delta}\|_2 = \mu\}$. Let $\lambda_\mu \in (0, (\sqrt{c} - 8M\mu\sqrt{s})/2]$. If the model error distribution satisfies the tail condition:*

$$\kappa_{\lambda_\mu} < \left(1 + \frac{64}{21}e^{-\frac{3}{2}}\right)^{-1}$$

*then for all $\boldsymbol{\Delta} \in K(S, \mu)$ it holds that*

$$L(\boldsymbol{\beta}^\star + \boldsymbol{\Delta}) - L(\boldsymbol{\beta}^\star) - \langle \nabla L(\boldsymbol{\beta}^\star), \boldsymbol{\Delta}\rangle \geq \kappa_1\|\boldsymbol{\Delta}\|_2\left(\|\boldsymbol{\Delta}\|_2 - \kappa_2\sqrt{\frac{\log p}{n}}\|\boldsymbol{\Delta}\|_1\right) \qquad (7)$$

*with probability at least $1 - \alpha_1 \exp(-\alpha_2 n)$, for some $\alpha_1, \alpha_2 > 0$, where $\kappa_1 = \frac{1}{4}\kappa_{RE}(C(1-\kappa_{\lambda_\mu}) - 2e^{-\frac{3}{2}})$ and $\kappa_2 = 97CM^2\sqrt{s}$, with $C = (21/32) + 2e^{-3/2}$.*

A proof is given in the appendix.

**Remarks.** Noting that for any $\boldsymbol{\Delta} \in C(S)$, $\|\boldsymbol{\Delta}\|_1 \leq 4\|\boldsymbol{\Delta}_S\|_1 \leq 4\sqrt{s}\|\boldsymbol{\Delta}\|_2$, the bound (7) implies that

$$L(\boldsymbol{\beta}^\star + \boldsymbol{\Delta}) - L(\boldsymbol{\beta}^\star) - \langle \nabla L(\boldsymbol{\beta}^\star), \boldsymbol{\Delta}\rangle \geq \frac{\kappa_1}{2}\|\boldsymbol{\Delta}\|_2^2$$

as long as $n > 64s\kappa_2^2 \log p$.

Lemma 3 indicates that the region of strong convexity is controlled by parameter $c$ via the condition $\|\boldsymbol{\Delta}\|_2 \leq \mu$ where $\mu < \sqrt{c}/(8\sqrt{s}M)$. The convexity of the loss rests on a condition related to the tail of the error distribution, which is required so that $\kappa_1 > 0$. We consider a specific example to illustrate that the requirements on the tail of the error distribution are very mild.

EXAMPLE. Let $\mu = c^{1/4}/(4\sqrt{s}M)$, $\lambda_\mu = \sqrt{c}/2 - c^{1/4}$. Hence $c$ must be chosen so that $P(|\eta_i| > \sqrt{c}/2 - c^{1/4}) \leq 0.59$. This translates into the conditions

$c > 2.42$ for the Laplace(0,1) distribution,

$c > 2.45$ for the Normal(0,1) distribution,

$c > 2.47$ for the Student's t distribution with 4 degrees of freedom, and

$c > 2.58$ for the Cauchy(0,1) distribution.

These conditions are quite similar. Nevertheless, except for the Laplace distribution, the heavier the tail, the larger the lower bound on $c$ needs to be in order to secure restricted strong convexity, which makes sense as the number of large outliers is expected to increase. It is important to note, however, that while a large value of $c$ extends the convexity region, it reduces the resilience to outliers (via the gradient bound). Thus the choice of $c$ is key in guaranteeing both fast convergence rates and robustness, as shall be made explicit in the next section.

We conclude this section by noting that under similar tail conditions on the error, the MD-Lasso loss function is (simply) convex with asymptotic probability 1 in the set $\mathcal{H}_c = \{\boldsymbol{\beta}^\star + \boldsymbol{\Delta} : \|\boldsymbol{\Delta}\|_1 \leq \sqrt{c}/M\}$. The proof is similar to that of Lemma 3 and is thus omitted.

## 4.3 Consistency results

We now state the results on the consistency and convergence rates for the MD-Lasso estimator. We present two theorems, both leveraging the restricted strong convexity and gradient boundedness properties. The first theorem builds on the gradient bound of Lemma 1, and is thus preferred if the variance of the error is undefined. The second theorem uses the gradient bound of Lemma 2 and is preferred for errors finite variance.

**Theorem 1** *Consider the linear regression model (1) and assume that the support of the true model coefficients $\boldsymbol{\beta}^\star$ has cardinality $s$. Let $\mathcal{H}_c := \{\boldsymbol{\beta}^\star + \boldsymbol{\Delta} : \|\boldsymbol{\Delta}\|_1 \leq \sqrt{c}/M\}$. Under Assumptions $[A1], [A2], [A3]$, for any $\gamma \leq \sqrt{c}$, with $c$ such that $\kappa_{\sqrt{c}/2} < (1 + (64/21)e^{-3/2})^{-1} < 0.6$ given the MD-Lasso estimator (5) with scaling parameter $c$ and regularization parameter $\lambda_n = 2\xi_{c,\gamma}\sqrt{\log(p)/n}$, where $\xi_{c,\gamma}^2 = M^2\kappa_1^{-2}\left[(1-2\kappa_\gamma)\gamma^2\exp(-\gamma^2/c) + 2c\kappa_\gamma/e\right]e^{1/c}$, any of the solutions in $\mathcal{H}_c$ (there is at least one such solution) satisfies*

$$\|\hat{\boldsymbol{\beta}}_{\lambda_n} - \boldsymbol{\beta}^\star\|_2 \leq \frac{4\xi_{c,\gamma}}{(C(1-\kappa_{\sqrt{c}/2}) - 2e^{-\frac{3}{2}})\kappa_{RE}}\sqrt{\frac{s\log p}{n}} \quad (8)$$

*with probability at least $1-\alpha_1\exp(-\alpha_2 n\lambda_n^2)$, where $C = 21/32 + 2e^{-3/2} \approx 1.1$, and positive constants $\alpha_1, \alpha_2 > 0$.*

A proof is given in the appendix.

**Theorem 2** *Consider the linear regression model (1) and assume that the support of the true model coefficients $\boldsymbol{\beta}^\star$ has cardinality $s$. Let $\mathcal{H}_c = \{\boldsymbol{\beta}^\star + \boldsymbol{\Delta} : \|\boldsymbol{\Delta}\|_1 \leq \sqrt{c}/M\}$. Under Assumptions $[A1], [A2], [A3]$, given the MD Lasso estimator (5) with scaling parameter $c$ such that $\kappa_{\sqrt{c}/2} < (1 + (64/21)e^{-3/2})^{-1} < 0.6$ and regularization parameter $\lambda_n = 2\zeta_c\sqrt{\log(p)/n}$, where $\zeta_c$ is given* by $\zeta_c^2 = 4M^2 \kappa_1^{-2} E[\eta_i^2 e^{-\eta_i^2/c}] e^{1/c}$, any of the solutions in $\mathcal{H}_c$ (there is at least one such solution) satisfies

$$\|\hat{\boldsymbol{\beta}}_{\lambda_n} - \boldsymbol{\beta}^\star\|_2 \leq \frac{4\zeta_c}{(C(1 - \kappa_{\sqrt{c}/2}) - 2e^{-\frac{3}{2}})\kappa_{RE}} \sqrt{\frac{s \log p}{n}} \qquad (9)$$

with probability at least

$$1 - \alpha_1 \exp\Big(-\alpha_2 n \lambda_n^2 \frac{1 - \alpha_3 \sqrt{c \log(p)/n}}{1 + \alpha_3 \sqrt{c \log(p)/n}}\Big),$$

where $C = 21/32 + 2e^{-3/2} \approx 1.1$, and $\alpha_1, \alpha_2, \alpha_3 > 0$ constants.

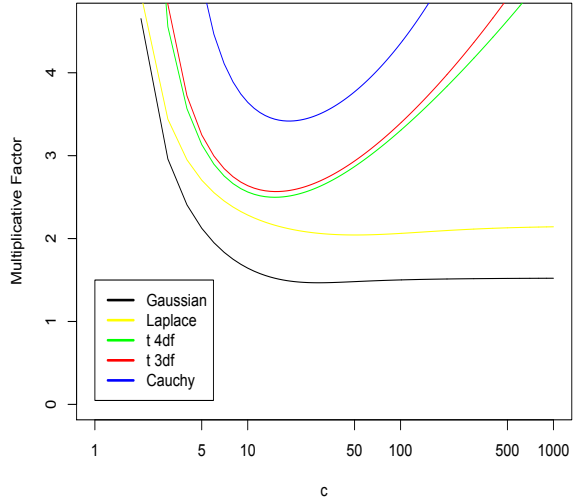

Figure 3: Scaling of the convergence rate for $\|\hat{\boldsymbol{\beta}} - \boldsymbol{\beta}^\star\|_2$ as a function of $c$

A proof is given in the appendix.

The bounds on $\|\hat{\boldsymbol{\beta}}_{\lambda_n} - \boldsymbol{\beta}^\star\|_2$ scale inversely with the restricted strong convexity constant of Lemma 3. This makes sense as the constant reflects the curvature of the loss function $L$ in a restricted set of directions around the true solution $\boldsymbol{\beta}^\star$: the higher the curvature the faster the convergence. On the other hand, the convergence rates and the regularization parameter $\lambda_n$ are proportional to the gradient bound of Lemmas 1 and 2. While strong convexity favors large values of $c$, the gradient bound favors small values, hence the "tension" between the two that we now elaborate upon.

Figure 3 depicts the multiplicative scaling factor in front of $\sqrt{s \log p/n}$ in the convergence rates of (8) and (9) for various error distributions. The convergence rates along with Figure 3 suggest the following points:

- Regardless of the error distribution, one should not set $c$ to values of much below a value of, say, 5. This makes sense intuitively, since a small $c$ means that we are not using many observations and operating on too small a sample size. Recall that as $c \to 0$ the estimator acts as a trimmed $\ell_1-$ penalized Least Squares estimator where all but one observations are trimmed out.

- As $c$ grows beyond an "optimum value"[2], the heavier the tail, the faster the multiplicative factor grows. This is aligned with the intuition that one should be more conservative the heavier the tail, and thus not set $c$ too large. Recalling that the MD-Lasso estimator is equivalent to the traditional Lasso estimator as $c \to \infty$, our results also corroborate the fact that the performance of the traditional Lasso estimator degrades dramatically in the presence of heavy-tailed noise.

- Interestingly, for lighter-tailed distribution (e.g. Laplace and Gaussian) the multiplicative factor flattens out and converges to a finite value as $c \to \infty$, provided that $c$ grows in a sample size dependent fashion so that $c \log(p)/n \to 0$. In particular, for sub-gaussian tails one recovers the results of the traditional Lasso estimator (up to a constant factor) as $c \to \infty$.

We also want to make a few remarks about global and local solutions. Recall that the MD-Lasso estimator is invex and locally convex but not globally convex. If the constraint region induced by the $\ell_1$-penalty resides within the local convexity region, the invexity of the loss is not compromised, every local minimum lies within the local convexity region and is also a global minimum. The results of Theorem 1 and Theorem 2 therefore apply to *any* solution of the MD-Lasso estimator. A sufficient condition for this case to hold is that $\sqrt{c} \geq M(\|\boldsymbol{\beta}^\star\|_1 + \bar{\lambda})$ where $\bar{\lambda}$ is the parameter corresponding to parameter $\lambda$ in the constrained version of the optimization. If the constraint region induced by the $\ell_1$-penalty merely intersects (or even resides outside of the local convexity region), local minima may exist outside of the convexity region. One might think that those local minima might be an issue, especially if no a-priori knowledge on the localization neighborhood of $\boldsymbol{\beta}^\star$ is available. However, we show in the next section that a simple composite gradient algorithm is able to remediate this.

## 5 Optimization

To solve (5) we first show a composite gradient method. A connection with re-weighted least squares will follow thereafter.

### 5.1 Composite gradient algorithm

We first show a composite gradient method. The need for solving very large problems has lead to a recent resurgence of interest in first-order optimization methods, such as the composite gradient method of Nesterov [27]. The algorithm proceeds with the following updates

$$\boldsymbol{\beta}^{(t+1)} = \arg\min_{\|\boldsymbol{\beta}\|_1 \leq b_0 \sqrt{s}} \left\{ L(\boldsymbol{\beta}^{(t)}) + \langle \nabla L(\boldsymbol{\beta}^{(t)}), \boldsymbol{\beta} - \boldsymbol{\beta}^{(t)} \rangle + \frac{\rho}{2}\|\boldsymbol{\beta} - \boldsymbol{\beta}^{(t)}\|_2^2 + \lambda \|\boldsymbol{\beta}\|_1 \right\}, \quad (10)$$

where the safety radius $b_0\sqrt{s}$ is imposed to ensure the good behavior of the algorithm in the first iterations and $b_0$ is chosen such that $\boldsymbol{\beta}^\star$ is feasible.

---

[2]Here we use the term "optimum" in a loose sense as our bounds may not be tight around the actual optimum.

**Instantiation of the composite gradient algorithm for MD-Lasso.** Denote by $S$ the *soft-thresholding operator* defined as

$$S_\lambda(u) = \text{sign}(u) \max(|u| - \lambda, 0), \tag{11}$$

where all operations are applied element-wise on a vector $u$. Each composite gradient update can be computed by (at most) two simple soft-thresholding operations. The first operation consists in computing $p_\rho(\boldsymbol{\beta}^{(t)}) = S_{\lambda_n/\rho}(\boldsymbol{\beta}^{(t)} - \frac{1}{\rho}\nabla L(\boldsymbol{\beta}^{(t)}))$. Let $r_i = y_i - \boldsymbol{X}_i \boldsymbol{\beta}^{(t)}$ denote the residual for the $i^{\text{th}}$ sample, and define

$$w_i = \frac{\exp\left(-\frac{1}{2c} r_i^2\right)}{\sum_{j=1}^n \exp\left(-\frac{1}{2c} r_j^2\right)}. \tag{12}$$

Let $\tilde{\boldsymbol{R}} = (\tilde{r}_1, \ldots, \tilde{r}_n)'$ where $\tilde{r}_i = r_i w_i$. Then $\nabla L(\boldsymbol{\beta}^{(t)}) = -\boldsymbol{X}'\tilde{\boldsymbol{R}}$, and $\tilde{\boldsymbol{R}}$ can be interpreted as a generalized residual. The thresholding operation boils down to the following simple step:

$$p_\rho(\beta^{(t)}) = S_{\lambda_n/\rho}(\boldsymbol{\beta}^{(t)} + \frac{1}{\rho}\boldsymbol{X}'\tilde{\boldsymbol{R}}). \tag{13}$$

If the $\ell_1$-norm of the projection exceeds the safety radius, a second soft thresholding operation has to be carried out to project onto the the $\ell_1$ ball of radius $b_0\sqrt{s}$ (see [1] for details.)

**Numerical convergence.** The following theorem guarantees that in the event where local minimizers exist, these reside within a ball of radius comparable to that of the consistent solutions of Theorems 1 and 2. The theorem is obtained by adapting the work of [1], noting that global convexity is actually not required for their results to hold (see the Appendix for more details).

**Theorem 3** *Assume that the design matrix $\boldsymbol{X}$ satisfies the restricted lower eigenvalue condition $\frac{1}{n}\|\boldsymbol{X}\boldsymbol{\Delta}\|_2^2 \leq \kappa_{RE^u}\|\boldsymbol{\Delta}\|_2^2$ for any $\boldsymbol{\Delta} \in C(S)$. Under the conditions of Theorem 1, for any of the solution $\hat{\boldsymbol{\beta}}$ in the local convexity region $\mathcal{H}_c$ of the MD-Lasso problem, there exists a contraction parameter $\delta \in (0,1)$ and constants $\alpha_1$ and $\alpha_2$ such that with asymptotic probability one the composite gradient iterates $\boldsymbol{\beta}^{(t)}$ satisfy*

$$\|\boldsymbol{\beta}^{(t)} - \hat{\boldsymbol{\beta}}\|_2^2 \leq \alpha_1 \|\hat{\boldsymbol{\beta}} - \boldsymbol{\beta}^\star\|_2^2$$

*for all*

$$t \geq \alpha_2 \frac{\log \frac{L(\boldsymbol{\beta}^{(0)}) - L(\hat{\boldsymbol{\beta}})}{\alpha_1 \|\hat{\boldsymbol{\beta}} - \boldsymbol{\beta}^\star\|_2^2}}{\log(1/\delta)}.$$

A proof is given in the appendix.

## 5.2 Re-weighted penalized least squares

Some interesting insights on the robustness of MD-Lasso can be gained by examining the descent direction in the composite gradient procedure, as explicated in (13). Given an initial solution $\boldsymbol{\beta}$, under the traditional squared loss one would get $\nabla_j L(\boldsymbol{\beta}) = -\frac{1}{n}\sum_{i=1}^n X_{i,j}(y_i - \boldsymbol{X}_i'\boldsymbol{\beta})$. For the MD-Lasso we have $\nabla_j L(\boldsymbol{\beta}) = -\sum_{i=1}^n w_i X_i^j(y_i - \boldsymbol{X}_i'\boldsymbol{\beta})$, where the weights $w_i$ are given in (12). Hence the descent direction for MD-Lasso can be seen as a "weighted version" of the direction for usual squared loss, where the weights $w_i$ can be interpreted as being proportional to the likelihood functions of individual data points, i.e., $w_i = \frac{\mathcal{L}(y_i|\boldsymbol{X}_i;\boldsymbol{\beta})}{\sum_{i=1}^{n}\mathcal{L}(y_i|\boldsymbol{X}_i;\beta)}$, where $\mathcal{L}$ denotes the likelihood function under Gaussian assumption. Thus data with high likelihood values are given more weights in the computation of the descent direction. Conversely, data with low likelihood values, which are more likely to be outliers, contribute less. The connection between the likelihood functions and weights provides an intuitive insight on the resilience of the original MD-Lasso to outliers.

We remark that a similar conclusion can be obtained by considering a first order approximation of the "log-sum-exp" term in (5) around an initial solution. This yields an approximate iterative procedure where given initial estimates, data are first re-weighted by $w_i$ in (12) and then passed to a traditional Lasso solver to provide new estimates, and the procedure is repeated until convergence. The following algorithm summarizes the procedure.

---

**Algorithm 1** Approximate MD-Lasso as Iterately Weighted Lasso

**Step 1**: Given initial estimate $\boldsymbol{\beta}^{(0)}$ for the regression coefficients (e.g. using ridge regression) compute weights $w_i = \frac{\exp(-\frac{1}{2c}r_i^2(\boldsymbol{\beta}^{(0)}))}{\sum_{i=1}^{n}\exp(-\frac{1}{2c}r_i^2(\boldsymbol{\beta}^{(0)}))}$, where $r_i(\boldsymbol{\beta}^{(0)}) = (y_i - \boldsymbol{X}_i\boldsymbol{\beta}^{(0)})$.

**Step 2**: Estimate $\boldsymbol{\beta}$ by minimizing

$$\hat{\boldsymbol{\beta}} = \arg\min_{\boldsymbol{\beta}} \sum_{i=1}^{n} w_i(y_i - X_i\boldsymbol{\beta})^2 + \lambda\|\boldsymbol{\beta}\|_1.$$

**Step 3**: If $\|\hat{\boldsymbol{\beta}} - \boldsymbol{\beta}^{(0)}\|^2 \leq \epsilon$ stop. Else, set $\boldsymbol{\beta}^{(0)} = \hat{\boldsymbol{\beta}}$ and go back to Step 1.

---

While the weighted least squares formulation illustrates most intuitively the robustness of the MD-Lasso loss, it requires running several individual Lasso problems. Even though the procedure can benefit from a warm start in each iteration, it is is computationally more intensive than running the composite gradient approach. Intuitively, the composite gradient approach can be interpreted as a "lazy update" version of the iterative weighted least squares.

# 6 Numerical Results

We compare the proposed MD-Lasso estimator with the LAD-Lasso [38], the Extended Lasso [28] and the traditional Lasso [33].

## 6.1 Simulation results

**Model setup.** We simulated data from the linear regression model

$$\boldsymbol{Y} = \boldsymbol{X}\boldsymbol{\beta}^\star + \boldsymbol{\eta}.$$

For the predictors, we consider two data generation models:

(P1) Toeplitz design: The $n \times p$ predictor matrices $\boldsymbol{X}$ have rows sampled independently from $\mathcal{N}(\boldsymbol{0}, \boldsymbol{\Sigma}_X)$ where $(\boldsymbol{\Sigma}_X)_i^j = 0.5^{|i-j|}$.

(P2) Factor model with two factors: let $\phi^1$ and $\phi^2$ be two latent variables following i.i.d. standard normal distributions. Each predictor variable $X^k$, for $k = 1, ,p$, is generated as $X^k = f^{k,1}\phi^1 + f^{k,2}\phi^2 + \epsilon^k$, where $f_k^1, f_k^2$ and $\epsilon^k$ have i.i.d. standard normal distributions for all $k = 1, \ldots, p$.

For the error term distribution, we consider five cases:

(E1) Normal: $\eta \sim \mathcal{N}(0,1)$.

(E2) Laplace: $\eta \sim \text{Laplace}(0,1)$.

(E3) Mixture of Gaussians: $\eta \sim \frac{h\mathcal{N}(0,1)+(1-h)\mathcal{N}(0,\sqrt{225})}{\sqrt{0.9*1+0.1*225}}$ where $h \sim \text{Bernoulli}(0.9)$.

(E4) Student's t with degrees of freedom 4: $\eta \sim \text{Student}(0,4)$.

(E5) Cauchy: $\eta \sim \text{Cauchy}(0,1)$.

In each simulation study, we consider both $n = 200$ and $n = 1000$ observations, and $p = 1000$ predictors. The entries of true model coefficient vector $\boldsymbol{\beta}^\star$ are set to be 0 everywhere, except for a randomly chosen subset of $s$ coefficients, which are chosen independently and uniformly in $(1, 3)$. The size $s$ of the set of non-zero coefficients is randomly set between 3 and 10.

**Parameter tuning.** We consider holdout-validated estimates, which are obtained by selecting the tuning parameter that minimizes the average loss on a validation set. We hold the parameter $c$ fixed at various values. The theoretical results suggested that a value of $c$ in the range 5-20 is a good tradeoff for most error distributions. The numerical results suggest that an optimal value is in the lower part of this range and a value of $c = 5$ will turn out to be a good default value for this parameter.

**Performance metrics.** To measure the estimation accuracy, we report the model error defined as

$$ME(\hat{\boldsymbol{\beta}}, \boldsymbol{\beta}^\star) = (\hat{\boldsymbol{\beta}} - \boldsymbol{\beta}^\star)'\boldsymbol{\Sigma}_X(\hat{\boldsymbol{\beta}} - \boldsymbol{\beta}^\star).$$

To measure variable selection accuracy, we use the $F_1$ score [35] defined by $F_1 = 2PR/(P + R)$, where $P$ is precision (fraction of correctly selected variables among selected variables) and $R$ is recall (fraction of correctly selected variables among true relevant variables).

**Results.** For each setting, we present the average of the performance measure based on 100 simulations. Figure 4 and Figure 5 provide boxplots for the Model Error and the variable selection accuracy, respectively. In the figures MD-x denotes MD-Lasso with $c = x$ and $x = 1, 2, 5, 10, 25, 50, 100$, Lasso denotes the Least Squares Lasso, LAD denotes the Least Absolute Deviation Lasso, and ExLasso denotes the Extended Lasso.

From the figures we can see that the simulations results are in agreement with the theoretical results of Section 4. Specifically if the scaling parameter is too small, the performance of the MD-Lasso method degrades, as the restricted strong convexity property is violated. As expected, the performance of MD-Lasso gets closer to that of Lasso as $c$ becomes large. For light tail distributions (e.g. Gaussian and Laplace) we see that as long as $c$ is larger or equal to the minimum value required for restricted strong convexity, the performance of MD-Lasso is quite insensitive to the choice of $c$, while the sensitivity increases for heavy tailed distributions (e.g. Student's t and Cauchy). We note that in each setting we verified that local minima are not an issue for MD-Lasso. To do so we applied the MD-Lasso method to the same problem instance 10 times, each time with a random starting point (we made sure that the starting points were sufficiently far apart so that some of them were clearly outside of the local convexity region), and measured the error between the various iterates and the solution output the first time. In agreement with Theorem 2, the iterates seemed to converge geometrically to the same fixed point.

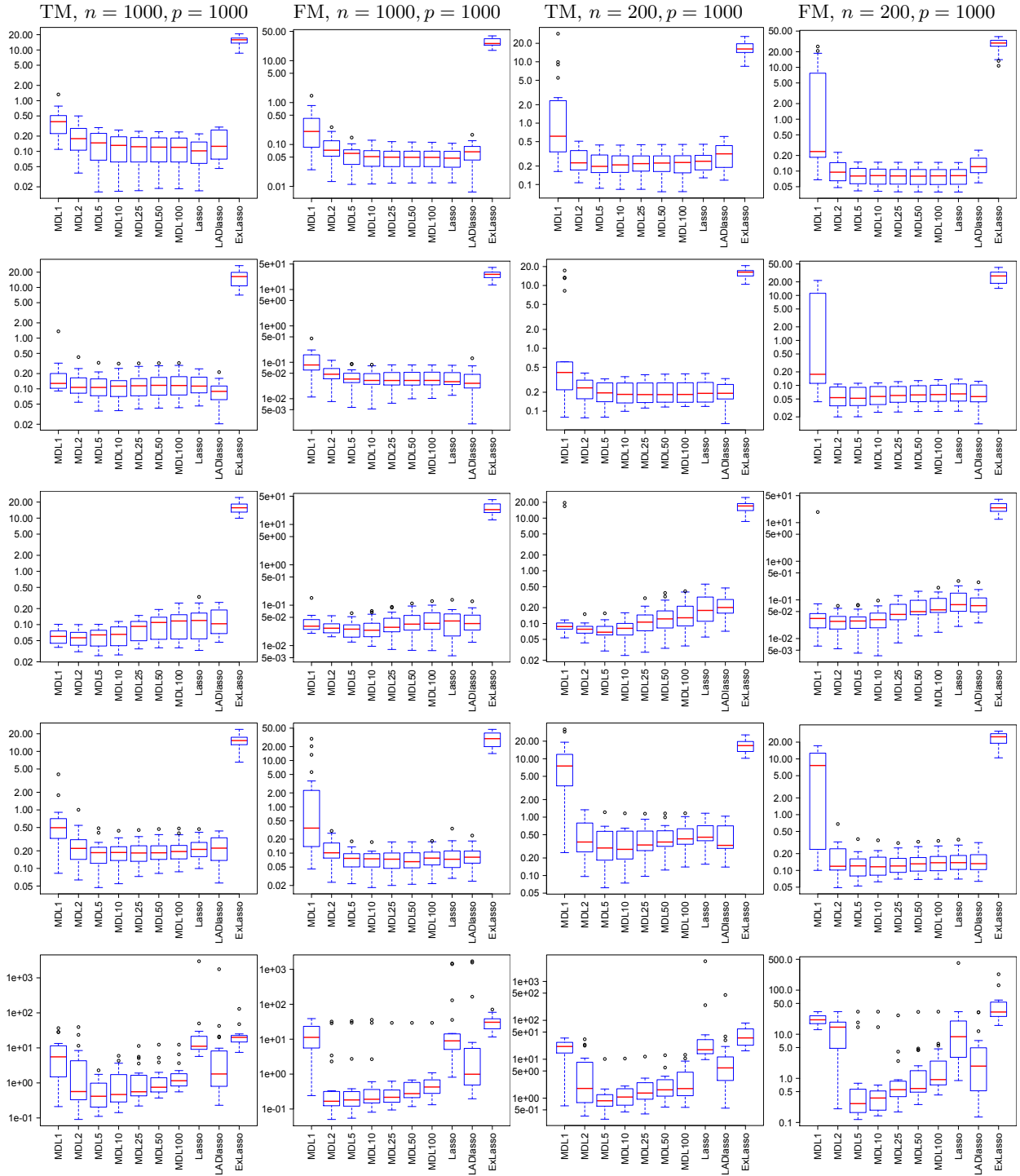

Figure 4: Model error (the lower the better) for the comparison methods and, from top to bottom row, errors with a Gaussian, Laplace, Gaussian Mixture, Student t (4 df) and Cauchy distribution. TM=Toeplitz Model, FM=Factor Model

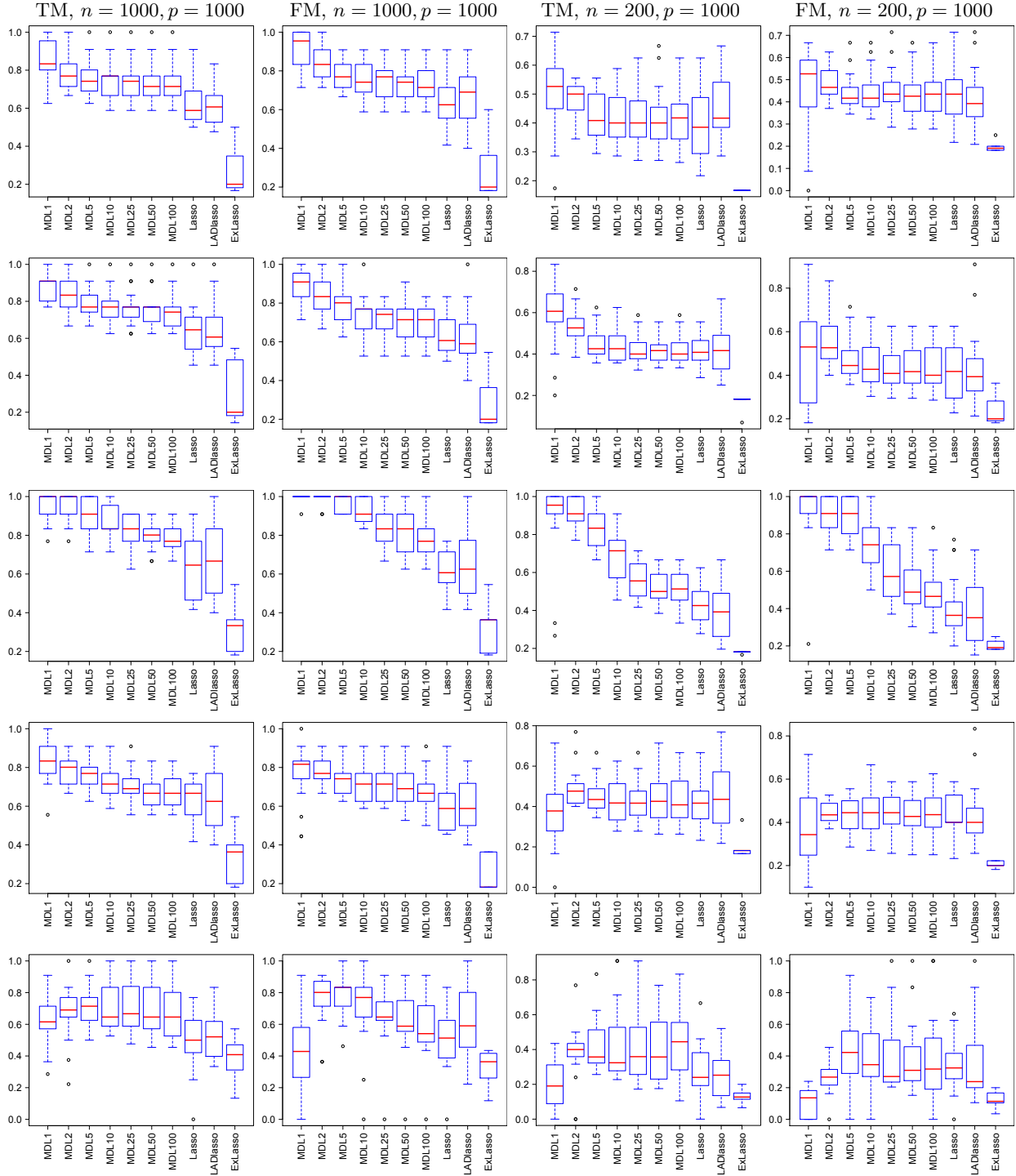

Figure 5: Variable selection accuracy (F1 score, the higher the better) for the comparison methods and, from top to bottom row, errors with a Gaussian, Laplace, Gaussian Mixture, Student t (4 df) and Cauchy distribution. TM=Toeplitz Model, FM=Factor Model

Remarkably the variable selection accuracy of MD-Lasso is generally superior to that of the other comparison methods, regardless of the error distribution. It is intriguing to note that in many cases the variable selection accuracy of MD-Lasso decreases monotonically with the scaling parameter, suggesting that the strong convexity of the loss might be more influential on the model error than on the variable selection accuracy. For Laplace distributed errors, LAD-Lasso performs the best in terms of model error. This can be explained by the fact that the noise distribution matches the LAD-Lasso loss exactly. However, MD-Lasso achieves higher variable selection accuracy. Interestingly even for Gaussian errors, MD-Lasso is able to outperform the Least Squares Lasso not only in terms of variable selection accuracy but also, in certain cases, in terms of model error (see the results for $n = 200, p = 1000$ and Toeplitz Model). The results for Cauchy distributed errors underscore the need for a non-convex loss functions as offered by MD-Lasso, and the limited ability of convex loss functions (including LAD-Lasso) in dealing with very large outliers.

## 6.2 Application to eQTL mapping

We apply MD-Lasso and other methods for comparison to the task of expression quantitative trait locus (eQTL) mapping. The main goal of eQTL studies is to identify the genetic variants (SNPs) that are associated with gene expression traits. In our analysis we use data on Alzheimers disease (AD) generated by Harvard Brain Tissue Resource Center and Merck Research Laboratories[3]. The dataset concerns $n = 206$ AD cases with SNPs and expression levels in the visual cortex. We study the associations between $p = 18137$ candidate SNPs and the expression levels of *APOE* gene, which is a key Alzeimer's gene [36]. Specifically, persons having an *APOE* $\epsilon 4$ allele have an increased chance of developing the disease; those who inherit two copies of the allele are at even greater risk.

The tuning parameters for all methods were chosen using a five-fold cross validation. We set $c = 5$ in MD-Lasso. To start, we investigated the Normal QQ-plots of the residuals from different regression methods and saw that the residuals from the fitted regressions have very heavy right tails. As an example the plot of the MD-Lasso is shown in Figure 6(a); the plots for the competing methods look similar. This suggests that robust methods are indeed recommended for this eQTL data analysis.

For ease of comparison, we first focus on a *cis*-eQTL analysis, namely we look into the subsets of SNPs within chromosome 19 (where gene *APOE* is located). To get a measure of confidence in the associations identified, we apply the bootstrap procedure (see [11] for a review) as follows. Given the original data, we randomly draw 100 datasets by sampling with replacement the rows of the original data, so that each dataset has the same number of rows as the original data. We then apply the comparison methods to each of the 100 bootstrap datasets. For each SNP selected using the original dataset, we count the number of times it appears in the bootstrap datasets. There is a sharp contrast among methods with respect to the number of selected SNPs. MD-Lasso and LAD-Lasso select fewer SNPs than Extended Lasso and Lasso (19 selected coefficients for MD-Lasso, 20 for LAD-Lasso, and over a 100 for Lasso and Extended Lasso). For each method, the top 20 selected SNPs according to the amplitude of their regression coefficients are listed in Table 1 along with their "confidence score". From Table 1 we can see that MD-Lasso and LAD-Lasso share 9 common SNPs. In contrast Extended Lasso share no common SNPs with MD-Lasso or LAD-Lasso, and Lasso shares at most 2 common SNPs with other methods. The SNPs identified by MD-Lasso are selected in average 71% of the time in the bootstrap datasets, those by LAD-Lasso 65% of the time, those by Extended Lasso 35% of the time, and those by Lasso 37% of the time. While a low variability in the selection process is not a guarantee that the selection includes the SNPs of interest, a high variability in the selection results (and a corresponding low confidence score like for Lasso and Extended Lasso) makes a consistent selection of interesting SNPs less plausible. We thus hypothesize that non-robust methods may select too many spurious associations due to their inability to cope with outliers and heavy-tailed errors.

---
[3]http://sage.fhcrc.org/downloads/downloads.php

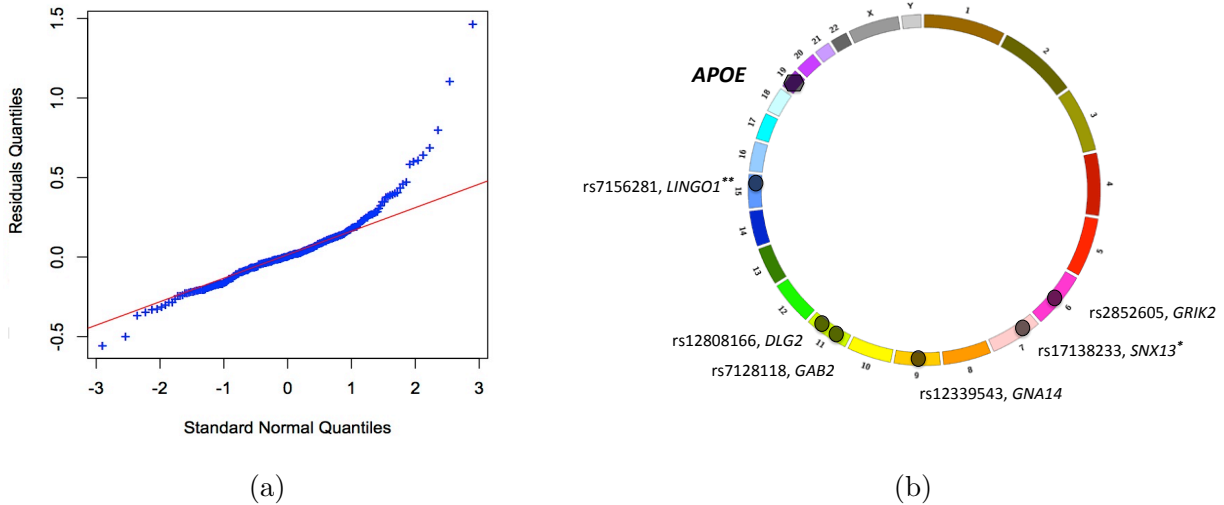

(a)  (b)

Figure 6: (a) Normal QQ-plots of residuals for MD-Lasso in the eQTL study. (b) "Circle" graph of the chromosomes highlighting the location of the target *APOE* gene, a subset of SNPs selected by MD-Lasso only (unmarked), by both MD-Lasso and LAD-Lasso (marked as *), and by Lasso only (marked as **) for the trans-eQTL study.

We now focus on the results obtained by the comparison methods on the full set of chromosomes. As the genetics of Alzheimer's disease are not yet fully understood, the variable selection results can only lead to qualitative statements about the performance of each method. To provide a more quantitative assessment, we evaluate the predictive accuracy of the various methods by randomly partitioning the data into training and test sets, using 150 observations for training and the remainder for testing. To get a sense of how a robust criterion performs, we also tested trimmed Lasso regression, removing the worst 10% observations according to the absolute residuals. We computed both the absolute prediction error and squared prediction error for the testing set for the model estimated using the training set. We repeated this process 20 times (using 20 random partitions). The results are presented in Table 2. Overall the predictive performance of MD-Lasso is superior to the other methods. We can also see that trimming is not as beneficial as using MD-Lasso or LAD-Lasso.

Table 1: Top 20 selected SNPs on chromosome 19 by comparison methods for the cis-eQTL study and their "confidence score" (the number of times the SNP is selected in 100 bootstrap samples).

| MD-Lasso | | LAD-Lasso | | Extended Lasso | | Lasso | |
|---|---|---|---|---|---|---|---|
| rs5021327 | 86 | rs1120559 | 81 | rs2285751 | 100 | rs280519 | 87 |
| rs280519 | 77 | rs1654322 | 75 | rs11882861 | 52 | rs2162296 | 71 |
| rs16964772 | 76 | rs11882490 | 74 | rs1433078 | 45 | rs10408465 | 69 |
| rs2112460 | 76 | rs3745297 | 74 | rs17314711 | 43 | rs12459372 | 67 |
| rs7249518 | 76 | rs2116877 | 70 | rs10409463 | 41 | rs16980543 | 63 |
| rs1599860 | 74 | rs3746006 | 68 | rs12460915 | 37 | rs12327600 | 42 |
| rs1673130 | 72 | rs353989 | 67 | rs2419549 | 37 | rs13730 | 40 |
| rs1120559 | 71 | rs10404242 | 66 | rs11665711 | 32 | rs1357879 | 39 |
| rs11672071 | 71 | rs11878850 | 66 | rs16973403 | 32 | rs1402325 | 35 |
| rs4805590 | 71 | rs16964772 | 65 | rs1402325 | 30 | rs17314711 | 28 |
| rs2395891 | 70 | rs3108549 | 63 | rs1422259 | 29 | rs1549951 | 17 |
| rs352826 | 69 | rs16964420 | 61 | rs12975977 | 28 | rs2304184 | 16 |
| rs7246997 | 69 | rs184239 | 59 | rs276731 | 28 | rs2395891 | 16 |
| rs10404242 | 67 | rs2292033 | 59 | rs1013414 | 27 | rs11084566 | 15 |
| rs3745297 | 67 | rs420703 | 59 | rs12459372 | 27 | rs16964772 | 13 |
| rs2301742 | 66 | rs1549951 | 58 | rs1560730 | 27 | rs1120559 | 12 |
| rs2304184 | 64 | rs16973403 | 58 | rs10414066 | 26 | rs11882490 | 12 |
| rs11084566 | 63 | rs2301742 | 58 | rs12609039 | 24 | rs10412301 | 11 |
| rs3108549 | 63 | rs280519 | 58 | rs2195948 | 22 | rs276725 | 11 |
| NA | NA | rs11881644 | 57 | rs12976494 | 21 | rs2304185 | 10 |

To conclude the eQTL analysis, we discuss some biologically interesting SNPs selected by various methods. These are depicted in Figure 6(b), which shows the chromosomes, highlights the position of the target *APOE* gene and the selected SNPs along with their closest gene. To facilitate the following discussion, we refer to the genes close to the corresponding SNPs. We first describe results pertaining to MD-Lasso, which are *not* found by other methods. Gene *DLG2* is a memory-associated protein known to be associated to Schizophrenia. However, a recent study showed that conservation of *DLG2* functions could potentially reduce the symptoms of Alzheimer's [19]. It has been shown that the inactivation of the gene *GRIK2* can cause severe learning disabilities. Gene *GNA14* has been identified by several studies as linked to Alzheimer's disease progression (see, e.g.

Table 2: Average test absolute error (MAE) and square error (MSE) with standard deviation for the models output by MD-Lasso and representative comparison methods on the eQTL dataset. (Smaller values indicate higher predictive accuracy).

|  | MD-Lasso | LAD-Lasso | Extended Lasso | Lasso | Trimmed Lasso |
|---|---|---|---|---|---|
| MAE | $11.90 \pm 1.52$ | $20.00 \pm 3.04$ | $175.75 \pm 14.25$ | $38.62 \pm 5.45$ | $35.71 \pm 4.46$ |
| MSE | $6.11 \pm 1.39$ | $12.18 \pm 2.99$ | $858.90 \pm 160.85$ | $20.13 \pm 9.87$ | $18.22 \pm 9.15$ |

[3]). It has been indicated that SNPs in gene *GAB2* can modify the risk of late-onset Alzheimer's disease in *APOE $\epsilon$4* carriers and plays an important role in Alzheimer's pathogenesis (see, e.g. [30]). Remarkably, MD-Lasso was the only method to select SNPs in the coding region of *GAB2*. We also checked for interesting SNPs selected by other methods. Most of them were also selected by MD-Lasso. For instance, SNP *rs17138233*, located within gene *SNX13*, was selected by *both* MD-Lasso and LAD-Lasso. The carboxyl terminal fragment of *SNX13* was reported to associate with activated *H-Ras* [14], which has been implicated in the process of neurodegeneration in Alzheimer's disease [4]. Our finding is quite intriguing as the functional consequence of the interaction between *SNX13* and *H-Ras* is not fully understood. An example among the few interesting SNPs discarded by MD-Lasso is *rs7156281*, located near gene *LINGO-1* , which was identified by Lasso only. *LINGO-1* is known to be involved in neurodegenerative processes including Alzheimer's disease [23].

Overall, our results suggest that the MD-Lasso method achieves greater predictive accuracy and stability than other methods, and is successful in identifying plausible and relevant SNPs in eQTL mapping.

# 7 Concluding Remarks

We have shown that combining minimum distance estimation with $\ell_1$-penalization is useful for robust high-dimensional regression. Our theoretical results indicate that the proposed MD-Lasso estimator can achieve optimal convergence rates even under heavy-tailed error distributions. These results hinge on the selection of a scaling parameter of MD-Lasso. If the scaling parameter is very large, MD-Lasso is identical to standard least-squares Lasso. We have shown that there are intermediate values of this parameter that lead, for a broad class of problems, to optimal convergence rates while still allowing a certain fraction of outliers in the data. Combining robustness with fast convergence rates requires non-convexity of the loss function, and the objective function can have multiple minima as a consequence. Our results hold for all local minima within a certain radius of the desired solution and we have provided reasonable conditions under which they hold for all local minima of the objective function. These desirable properties were confirmed by numerical examples. The MD-Lasso framework should prove equally useful in other statistical models such as generalized linear models, which will be investigated in a future study.

# Appendix

## A  Proof of Lemma 1 and Lemma 2

Define $p_i$ for $i = 1, \ldots, n$ as

$$p_i = \frac{\exp(-\frac{\eta_i^2}{2c})}{\sum_{j=1}^n \exp(-\frac{\eta_j^2}{2c})}$$

and $f^j(\eta_1, \ldots, \eta_n) = \sum_{i=1}^n \eta_i p_i X_i^j$. Note that $X_i^j$ are constants with $|X_i^j| \leq M$. We have $\nabla_j L(\beta^\star) = f^j(\eta_1, \ldots, \eta_n)$.

**Notation and lemma.** Let the event $E_{\gamma,\lambda}^-$ for $\lambda \geq 0$ and $\gamma < 1$ be defined as $\sum_{i=1}^n 1\{|\eta_i| < \lambda\} < n\gamma(1 - \kappa_\lambda)$. With Hoeffding's inequality we have for $\gamma < 1$

$$P(E_{\lambda,\gamma}^-) \leq \exp\left(-2n(1-\gamma)^2(1-\kappa_\lambda)^2\right). \tag{14}$$

Likewise let $E_{\gamma,\lambda}^+$ be defined as $\sum_{i=1}^n 1\{|\eta_i| \geq \lambda\} > n\gamma\kappa_\lambda$. With Hoeffding's inequality we have for $\gamma > 1$ again

$$P(E_{\lambda,\gamma}^+) \leq \exp(-2n(\gamma-1)^2\kappa_\lambda^2). \tag{15}$$

**Showing that $E[f^j] = 0$.** If $E[\eta]$ is well defined this is straightforward (e.g. for Gaussian, Laplace, Student's, Gaussian mixture etc). The following deals with the case where $E[\eta]$ is undefined (e.g. Cauchy). Assume that the error term has a symmetrical and bounded probability density function $\mu$. Namely for all $x \in \mathbb{R}$, $\mu(x) = \mu(-x)$ and there exists an $A > 0$ such that $0 \leq \mu(x) \leq A < \infty$.

Since the predictors are bounded, we have by Hölder's inequality

$$|E[f^j]| = |\sum_{i=1}^n E[\eta_i p_i] X_i^j| \leq \|X^j\|_\infty \sum_{i=1}^n |E[\eta_i p_i]| = M \sum_{i=1}^n |E[\eta_i p_i]|. \tag{16}$$

Now

$$|E[\eta_i p_i | \eta_j \text{ for all } j \neq i]| \leq \frac{1}{b} |E[\eta_i \exp\left(-\frac{\eta_i^2}{2c}\right)]|, \tag{17}$$

where $b = \sum_{j \neq i} \exp(-\frac{\eta_j^2}{2c})$. We have

$$\begin{aligned} E[\eta_i \exp\left(-\frac{\eta_i^2}{2c}\right)] &= \int_{-\infty}^\infty \mu(x) x \exp\left(-\frac{x^2}{2c}\right) dx \\ &= -\int_0^\infty \mu(x) x \exp\left(-\frac{x^2}{2c}\right) dx + \int_0^\infty \mu(x) x \exp\left(-\frac{x^2}{2c}\right) dx, \end{aligned}$$

where the last equation comes from the fact that $\mu$ is symmetrical.

Now we have

$$0 \leq \int_0^\infty \mu(x) x \exp\left(-\frac{x^2}{2c}\right) dx \leq A \int_0^\infty x \exp\left(-\frac{x^2}{2c}\right) dx = Ac$$

Hence $\int_0^\infty \mu(x) x \exp\left(-x^2/(2c)\right) dx$ is finite and thus $E[\eta_i \exp\left(-\eta_i^2/(2c)\right)]| = 0$. Together with (16) and (17) we conclude that $E[f^j] = 0$.

**McDiarmid's inequality.** We have to find bounds $\delta_i$ such that for all $\eta_1, \ldots, \eta_n, \tilde{\eta}_i$ and all $i$,

$$|f^j(\eta_1, \ldots, \eta_i, \ldots, \eta_n) - f^j(\eta_1, \ldots, \tilde{\eta}_i, \ldots, \eta_n)| \leq \delta_i. \tag{18}$$

Once we have these, since $E[f] = 0$, by McDiarmid's inequality we have for all $t > 0$

$$P(|f| > t) \leq 2 \exp\left(-\frac{2t^2}{\sum_i \delta_i^2}\right) \tag{19}$$

If the bounds $\delta_i$ in (18) hold only with probability $1 - \alpha$, we have

$$P(|f| > t) \leq 2 \exp\left(-\frac{2t^2}{\sum_i \delta_i^2}\right) + \alpha \qquad (20)$$

**Bounds.** Now, for a given $\eta_1, \ldots, \eta_{i-1}, \eta_{i+1}, \ldots, \eta_n$, we have, if $\eta_i$ can take any real value,

$$\max_x |f^j(\eta_1, \ldots, \eta_{i-1}, x, \eta_{i+1}, \ldots, \eta_n)| \leq M \max_{x \geq 0} \left|\frac{x \exp(-\frac{x^2}{2c})}{b + \exp(-\frac{x^2}{2c})}\right| + \text{const}$$

$$\leq M \max_{x \geq 0} \left|\frac{x \exp(-\frac{x^2}{2c})}{b}\right| + \text{const},$$

where $b = \sum_{j \neq i} \exp(-\frac{\eta_j^2}{c})$. Now, $\max_{x \geq 0} x \exp(-\frac{x^2}{2c})$ is attained at $x = \sqrt{c}$ and the maximal value is $\sqrt{c/e}$. Hence the lhs in (18) is bounded by

$$\max_{\eta, \tilde{\eta}} |f^j(\eta_1, \ldots, \eta_i, \ldots, \eta_n) - f^j(\eta_1, \ldots, \tilde{\eta}_i, \ldots, \eta_n)| \leq 2M \max_x |f^j(\eta_1, \ldots, \eta_{i-1}, x, \eta_{i+1}, \ldots, \eta_n)|$$

$$\leq 2M \frac{\sqrt{c/e}}{b}.$$

On the other hand, if $|\eta|$ is constrained to be within $\lambda$, and $\lambda \leq \sqrt{c}$ then the equivalent bound becomes

$$2 \max_{x:|x| \leq \lambda} |f^j(\eta_1, \ldots, \eta_{i-1}, x, \eta_{i+1}, \ldots, \eta_n)| \leq \frac{2\lambda \exp(-\frac{\lambda^2}{2c})}{b}. \qquad (21)$$

Fixing a particular value of $\lambda$, we have by (15) that with probability at most $\exp(-2n\kappa_\lambda^2)$ a proportion of at least $2\kappa_\lambda$ of all samples have value larger in absolute value than $\lambda$. Hence the sum of all $\delta_i^2$ is bounded with probability $1 - \exp(-2n\kappa_\lambda^2)$ by

$$n^{-1} \sum_{i=1}^n \delta_i^2 \leq M(1 - 2\kappa_\lambda)\left(2\frac{\lambda \exp(-\frac{\lambda^2}{2c})}{b}\right)^2 + M(2\kappa_\lambda)\left(2\frac{\sqrt{c/e}}{b}\right)^2.$$

**Bounding $b$.** Let $b = \sum_{j \neq i} \exp(-\eta_j^2/(2c))$. Let $\mathcal{E}_{b,\lambda',\gamma}$ be the event

$$\sum_{j \neq i} 1\{|\eta_j| \geq \lambda'\} > n\gamma \kappa_{\lambda'}.$$

We have $E[\sum_{j \neq i} 1\{|\eta_j| \geq \lambda'\}] = (n-1)\kappa_{\lambda'}$. By Hoeffding's inequality,

$$P\left(\sum_{j \neq i} 1\{|\eta_j| \geq \lambda'\} > E[\sum_{j \neq i} 1\{|\eta_j| \geq \lambda'\}] + t\right) \leq \exp\left(\frac{-2t^2}{\sum_{j \neq i}(1-0)^2}\right),$$

since $1\{|\eta_j| \geq \lambda'\} \in [0, 1]$. Setting $t = (n(\gamma - 1) + 1)\kappa_{\lambda'}$ we obtain

$$P(\mathcal{E}_{b,\lambda',\gamma}) \leq \exp\left(\frac{-2((n(\gamma - 1) + 1)\kappa_{\lambda'})^2}{n - 1}\right)$$

and hence, since $\exp(-x)$ is decreasing,

$$P\Big(b < n\gamma\kappa_{\lambda'}\exp\big(-\frac{\lambda'^2}{2c}\big)\Big) \leq \exp\Big(\frac{-2\left(n(\gamma-1)+1\right)^2\kappa_{\lambda'}^2}{n-1}\Big).$$

For $\gamma = 2$ and $\lambda' = 1$ it follows that

$$P\Big(b \geq 2n\kappa_1\exp(-\frac{1}{2c})\Big) > 1 - \exp\Big(\frac{-2\left(n+1\right)^2\kappa_1^2}{n-1}\Big).$$

Hence, with probability at least $1 - \exp(-2n\kappa_{\sqrt{1}}^2) - \exp(-2n\kappa_\lambda^2)$,

$$\sum_{i=1}^n \delta_i^2 \leq \frac{M^2 e^{1/c}}{n\kappa_1^2}\Big[(1-2\kappa_\lambda)\lambda^2 e^{-\lambda^2/c} + \frac{2c\kappa_\lambda}{e}\Big].$$

In summary, we have from (20) that

$$P(|f^j| > t) \leq 2\exp\big(-2nb_{\lambda,c}t^2\big) + \exp(-2n\kappa_1^2) + \exp(-2n\kappa_\lambda^2), \tag{22}$$

where $b_{\lambda,c} = \frac{\kappa_1^2}{M^2 e^{1/c}}\Big[(1-2\kappa_\lambda)\lambda^2 + \frac{2c\kappa_\lambda}{e}\Big]^{-1}$ with $\lambda \leq \sqrt{c}$ By a union bound over the predictors we obtain

$$P(\max_{j=1,\ldots,p} |f_j(\eta_1,\ldots,\eta_n)| > t) \leq 2\exp\big(-2nb_{\lambda,c}t^2 + \log p\big)$$
$$+ \exp(-n\kappa_1^2/2 + \log p) + \exp(-2n\kappa_\lambda^2 + \log p),$$

We can set $t^2 \geq b_{\lambda,c}^{-1}\log(p)/n$ and $\lambda = \sqrt{c}$. We remarks that it seems safe to assume that $\log(p)/n \to 0$. However, if we choose $c$ as a function of $n$, we still need to make sure that $\kappa_\lambda = \omega(1/\sqrt{n})$, where $\lambda < \sqrt{c}$.

**Bernstein's inequality.** The random variable $z_i = \eta_i\exp(-\eta_i^2/(2c))$ is a mean-zero random variable. Furthermore it is bounded in absolute value by $\sqrt{c/e}$ and hence its variance is guaranteed to exist and is at most $c/e$ (regardless of whether or not the variance of $\eta_i$ is well-defined). We can thus apply Bernstein's inequality as follows.

$$P\Big\{\frac{\sum_{i=1}^n z_i}{n} > t\Big\} \leq \exp\Big(-\frac{n^2 t^2}{2(\sum_{i=1}^n E[z_i^2] + \sqrt{\frac{c}{e}}\frac{nt}{3})}\Big).$$

It also holds that

$$P\Big\{\frac{\sum_{i=1}^n z_i}{n} < -t\Big\} \leq \exp\Big(-\frac{n^2 t^2}{2\sum_{i=1}^n E[z_i^2]}\Big).$$

We now specialize the above bounds in cases where the variance of $z_i$ is computable in closed form.

**Gaussian errors.** If $(\eta_i)$ is a zero-mean gaussian sequence with variance $\sigma^2$, let $d_{\sigma,c} = (c/(2\sigma^2 + c))^{3/2}\sigma^2$. Then $E[z_i^2] = d_{\sigma,c}$. By a union bound over the predictors we obtain

$$P(\max_{j=1,\ldots,p} |f_j(\eta_1,\ldots,\eta_n)| > t)$$
$$\leq \exp\Big(-\frac{n}{2d_{\sigma,c} + 2\sqrt{c/e}(t/3)}t^2 + \log p\Big) + \exp\Big(-\frac{n}{2d_{\sigma,c}}t^2 + \log p\Big) + \exp(-n\kappa_1^2/2 + \log p).$$

If $c$ is finite, we can set $t^2 = 4\log(p)d_{\sigma,c}/n$. (This is enough because the term $2\sqrt{c/e}(t/3)$ becomes negligible compared to $d_{\sigma,c}$ as $n \to \infty$ as long as $\log(p)/n \to 0$.) If $c \to \infty$ while $c\log(p)/n \to 0$, we recover condition for the traditional Lasso namely:

$$t^2 = 4\sigma^2 \frac{\log p}{n}.$$

Note that if $c < \sigma^2$ we obtain $t^2 \sim 4\frac{\log p}{n}c$ which is similar (w.r.t. dependence in $c$) to what we got with McDiarmid's inequality.

**Laplacian errors.** If $(\eta_i)$ is a sequence of zero-mean Laplace$(0, 2b)$ random variables then

$$E[z_i^2] = -\frac{c^2}{4b^2} + \frac{\sqrt{2\pi}}{b}\Big(\frac{c}{2}\Big)^{3/2} e^{\frac{c}{4b^2}} \Big(1 + \frac{c}{2b^2}\Big) \bar{F}\Big(\frac{1}{b}\sqrt{\frac{c}{2}}\Big),$$

where $\bar{F}(\cdot)$ denotes the tail probability function of the standard normal distribution. Note that as $c \to \infty$, $E[z_i^2] \to 2b^2$. By a union bound over the predictors,

$$P(\max_{j=1,\ldots,p} |f_j(\eta_1,\ldots,\eta_n)| > t)$$
$$\leq \exp\Big(-\frac{n}{2d_{b,c} + 2\sqrt{c/e}(t/3)}t^2 + \log p\Big) + \exp\Big(-\frac{n}{2d_{b,c}}t^2 + \log p\Big) + \exp(-n\kappa_1^2/2 + \log p),$$

where $d_{b,c} = E[z_i^2]$. If $c$ is finite, we can set $t^2 = 4\log(p)d_{b,c}/n$. If $c \to \infty$ while $c\log(p)/n \to 0$, we obtain the condition

$$t^2 = 8b^2 \log(p)/n.$$

**Error distributions with undefined variances (e.g. Cauchy).** If the error distribution has an undefined variance, there is no hope of getting a gradient condition which would guarantee that the gradient is still finite as $c \to \infty$. While Bernstein inequality is still applicable, as we know that $E[z_i^2] \leq \frac{c}{e}$, McDiarmid's inequality yields tighter bounds in this case.

## B  Proof of Lemma 3

**Notation.** Let $r_i, i = 1,\ldots,n$ be the residual for $\boldsymbol{\beta} = \boldsymbol{\beta}^\star + t\boldsymbol{\Delta}$, . Then $r_i = (\eta_i - \langle \boldsymbol{X}_i, t\boldsymbol{\Delta}\rangle)$. Define $\tilde{p}_i$ for $i = 1,\ldots,n$ as

$$\tilde{p}_i = \frac{\exp(-\frac{r_i^2}{2c})}{\sum_{j=1}^n \exp(-\frac{r_j^2}{2c})}. \tag{23}$$

For any $\lambda \geq 0$ let $\kappa_\lambda = P(|\eta_i| > \lambda)$.

**Preliminary Lemma.** The following technical lemma establishes the conditions required on the Hessian to guarantee strong convexity of the loss $L$ in a neighborhood of the true parameter $\boldsymbol{\beta}^\star$.

**Lemma 4** *Let $\boldsymbol{A} \subset \mathbb{R}^p$ be star-shaped with respect to $\boldsymbol{\beta}^\star \in \mathbb{R}^p$, namely for any $\boldsymbol{\beta} \in \boldsymbol{A}$ and $t \in [0,1]$ it holds that $t\boldsymbol{\beta}^\star + (1-t)\boldsymbol{\beta} \in \boldsymbol{A}$. Let $f : \boldsymbol{A} \to \mathbb{R}$ be a twice-differentiable function. Let the second derivative of $f$ satisfy $\nabla^2 f(\boldsymbol{\beta}^\star + t\boldsymbol{\Delta})(\boldsymbol{\Delta}, \boldsymbol{\Delta}) > 2\kappa_1 \|\boldsymbol{\Delta}\|_2(\|\boldsymbol{\Delta}\|_2 - \kappa_2 \|\boldsymbol{\Delta}\|_1)$ for all $\boldsymbol{\Delta}$ such that $\boldsymbol{\beta}^\star + \boldsymbol{\Delta} \in \boldsymbol{A}' \subset \boldsymbol{A}$ and $t$ in $(0,1]$. Then for all $\boldsymbol{\Delta}$ such that $\boldsymbol{\beta}^\star + \boldsymbol{\Delta} \in \boldsymbol{A}' \subset \boldsymbol{A}$ we have $f(\boldsymbol{\beta}^\star + \boldsymbol{\Delta}) - f(\boldsymbol{\beta}^\star) - \langle \nabla f(\boldsymbol{\beta}^\star), \boldsymbol{\Delta} \rangle > \kappa_1 \|\boldsymbol{\Delta}\|_2(\|\boldsymbol{\Delta}\|_2 - \kappa_2 \|\boldsymbol{\Delta}\|_1)$.*

*Proof of Lemma 4:* Consider the function $g(t) := f(\boldsymbol{\beta}^\star + t\boldsymbol{\Delta}) - f(\boldsymbol{\beta}^\star) - \langle \nabla f(\boldsymbol{\beta}^\star), t\boldsymbol{\Delta} \rangle - t^2(\kappa_1 \|\boldsymbol{\Delta}\|_2(\|\boldsymbol{\Delta}\|_2 - \kappa_2 \|\boldsymbol{\Delta}\|_1))$. We need to show that $g(1) > 0$. It holds that $g(0) = g'(0) = 0$. Moreover $g''(t) = \nabla^2 f(\boldsymbol{\beta}^\star + t\boldsymbol{\Delta})(\boldsymbol{\Delta}, \boldsymbol{\Delta}) - 2(\kappa_1 \|\boldsymbol{\Delta}\|_2(\|\boldsymbol{\Delta}\|_2 - \kappa_2 \|\boldsymbol{\Delta}\|_1)) > 0$ on $(0,1]$. This implies that $g'$ is positive on $(0,1]$ and $g(1) > 0$. □

Lemma 4 indicates that it suffices to focus on the Hessian of $L$ and establish that a lower bound of the form $\nabla^2 L(\boldsymbol{\beta}^\star + t\boldsymbol{\Delta})(\boldsymbol{\Delta}, \boldsymbol{\Delta}) \geq 2\kappa_1 \|\boldsymbol{\Delta}\|_2(\|\boldsymbol{\Delta}\|_2 - \kappa_2 \|\boldsymbol{\Delta}\|_1)$ holds for all $\boldsymbol{\Delta}$ in $K(S, \mu)$ and $t \in (0,1]$.

**Condition on the Hessian of $L$.** We first provide the expression for $\nabla^2 L(\boldsymbol{\beta}^\star + t\boldsymbol{\Delta})(\boldsymbol{\Delta}, \boldsymbol{\Delta})$, for which we then provide a lower-bound. Let $\boldsymbol{\Delta} \in \mathbb{R}^p$. The gradient of $L$ evaluated at a vector $\boldsymbol{\beta} = \boldsymbol{\beta}^\star + t\boldsymbol{\Delta}$, $0 \leq t \leq 1$ can be expressed as

$$\nabla_j(L(\boldsymbol{\beta})) = -\sum_{i=1}^n \tilde{p}_i r_i X_i^j, \quad j = 1, \ldots, p, \tag{24}$$

Differentiating a second time we obtain

$$\nabla^2 L(\boldsymbol{\beta})(\boldsymbol{\Delta}, \boldsymbol{\Delta}) = \sum_{i=1}^n \left( \tilde{p}_i \left(1 - \frac{1}{c} r_i^2\right) s_i^2 \right) + \frac{1}{c} \left( \sum_{i=1}^n \tilde{p}_i r_i s_i \right)^2, \tag{25}$$

where $s_i = \boldsymbol{X}_i' \boldsymbol{\Delta}$.

Let $z_i = r_i^2/c$. We wish to lower-bound $f(z_i) = e^{-z_i/2}(1 - z_i)$. Let $a^2 < 1$. If $z_i \leq a^2$, it holds that $f(z_i) \geq (1-a^2)(1 - \frac{a^2}{2})$, noting that $\exp(-z_i/2) \geq (1 - \frac{z_i}{2})$. On the other hand if $z_i > a^2$ then $f(z_i) \geq -2e^{-\frac{3}{2}}$. Let

$$\psi(z) = \begin{cases} (1-a^2)(1-\frac{a^2}{2}) & \text{if } z \leq a^2 \\ -2e^{\frac{-3}{2}} & \text{if } z > a^2 \end{cases} \tag{26}$$

Then (25) can be lower-bounded as follows

$$\nabla^2 L(\boldsymbol{\beta}^\star + t\boldsymbol{\Delta})(\boldsymbol{\Delta}, \boldsymbol{\Delta}) \geq \frac{1}{\sum_{i=1}^n \exp\left(-r_i^2/(2c)\right)} \sum_{i=1}^n \psi(z_i) s_i^2.$$

**Goal.** In what follows we shall consider $K_\nu(S, \mu) = \{\boldsymbol{\Delta} \in C(S) : \|\boldsymbol{\Delta}\|_1 = \nu, \|\boldsymbol{\Delta}\|_2 = \mu\}$ and show that the probability of the event

$$\mathcal{E}(\nu) = \{\frac{1}{n} \sum_{i=1}^{n} \psi(z_i) s_i^2 < g(\nu, \mu), \text{ for some } \boldsymbol{\Delta} \in K_\nu(S, \mu)\}$$

is very small, where $g(\nu, \mu)$ shall be specified. Then we shall appeal to a peeling argument (see [29]) to prove that the event over all $\boldsymbol{\Delta} \in K(S, \mu) = \{\boldsymbol{\Delta} \in C(S) : \|\boldsymbol{\Delta}\|_2 = \mu\}$ is also very small.

We begin by proving a tail bound for

$$\tilde{Z}(\nu) = \sup_{\boldsymbol{\Delta} \in K_\nu(S,\mu)} \left| \frac{1}{n} \sum_{i=1}^{n} (\psi(z_i) - E(\psi(z_i))) s_i^2 \right|.$$

**Step 1: Lower-bounding $\frac{1}{n} \sum_{i=1}^{n} E\psi(z_i) s_i^2$.**

Let $\epsilon_n \in \mathbb{R}$ be such that $4M\sqrt{s}\mu/\sqrt{c} \leq \epsilon_n < a$ and let $\lambda_\mu = \sqrt{c}(a - \epsilon_n)$. We first show that

$$E[1\{z_i \geq a^2\}] = P(z_i \geq a^2) \leq \kappa_{\lambda_\mu},$$

where $\kappa_{\lambda_\mu} = P(|\eta_i| > \lambda_\mu)$.

Indeed, $P(z_i \geq a^2) = P(|\eta_i - tX_i'\boldsymbol{\Delta}| > a\sqrt{c}) = P(\eta_i > a\sqrt{c} + tX_i'\boldsymbol{\Delta}) + P(\eta_i < -a\sqrt{c} + tX_i'\boldsymbol{\Delta})$. By Hölder's inequality we have

$$|X_i'\boldsymbol{\Delta}| \leq \|X_i\|_\infty \|\boldsymbol{\Delta}\|_1 \leq 4M\sqrt{s}\|\boldsymbol{\Delta}\|_2 \leq \sqrt{c}\epsilon_n.$$

Now $P(\eta_i > a\sqrt{c} + tX_i\boldsymbol{\Delta}) \leq P(\eta_i > a\sqrt{c} - \sqrt{c}\epsilon_n)$ and $P(\eta_i < -a\sqrt{c} + tX_i\boldsymbol{\Delta}) \leq P(\eta_i < -a\sqrt{c} + \sqrt{c}\epsilon_n)$ Thus

$$P(|\eta_i - tX_i'\boldsymbol{\Delta}| > a\sqrt{c}) \leq P(|\eta_i| > \sqrt{c}(a - \epsilon_n)) = P(|\eta_i| > \lambda_\mu) = \kappa_{\lambda_\mu},$$

as desired.

We also have $E[1\{z_i < a^2\}] = 1 - P(z_i \geq a^2) \geq 1 - \kappa_{\lambda_\mu}$. Thus we obtain

$$E[\psi(z_i)] = (1 - a^2)(1 - \frac{a^2}{2}) P(z_i \leq a^2) - 2e^{-\frac{3}{2}} P(z \geq a^2) \tag{27}$$

$$\geq (1 - a^2)(1 - \frac{a^2}{2}) - \kappa_{\lambda_\mu} \left( (1 - a^2)(1 - \frac{a^2}{2}) + 2e^{-\frac{3}{2}} \right). \tag{28}$$

Hence if

$$\kappa_{\lambda_\mu} \leq \frac{(1 - a^2)(1 - \frac{a^2}{2})}{(1 - a^2)(1 - \frac{a^2}{2}) + 2e^{-\frac{3}{2}}},$$

we have $E[\psi(z_i)] \geq 0$. Then, noting that $\sum_{i=1}^{n} \exp(-r_i^2/(2c)) \leq n$, we obtain

$$E[\nabla^2 L(\beta^\star + t\boldsymbol{\Delta})(\boldsymbol{\Delta}, \boldsymbol{\Delta})] \geq \frac{1}{n} \sum_{i=1}^{n} E\psi(z_i) s_i^2 \tag{29}$$

$$\geq \frac{1}{n} \left( C_a(1 - \kappa_{\lambda_\mu}) - 2e^{-\frac{3}{2}} \right) \sum_{i=1}^{n} s_i^2$$

$$\geq \kappa_{RE} \mu^2 (C_a(1 - \kappa_{\lambda_\mu}) - 2e^{-\frac{3}{2}}), \tag{30}$$

where $C_a = (1 - a^2)(1 - \frac{a^2}{2}) + 2e^{-\frac{2}{3}}$. Here the second inequality comes from (28) and the last inequality is due to the Restricted Eigenvalue condition (6).

**Step 2: Upper-bounding $\tilde{Z}(\nu)$.**
We first show that
$$|\psi(z_i) - E\psi(z_i)| \leq \left[(1-a)^2(1-\frac{a^2}{2}) + 2e^{-\frac{3}{2}}\right].$$

Indeed if $\psi(z_i) \geq E\psi(z_i)$ we have

$$
\begin{aligned}
|\psi(z_i) - E\psi(z_i)| &= \psi(z_i) - E\psi(z_i) \\
&\leq (1-a^2)(1-\frac{a^2}{2}) - (1-a^2)(1-\frac{a^2}{2}) + \kappa_{\lambda_\mu}\left((1-a^2)(1-\frac{a^2}{2}) + 2e^{-\frac{3}{2}}\right) \\
&= \kappa_{\lambda_\mu}\left((1-a^2)(1-\frac{a^2}{2}) + 2e^{-\frac{3}{2}}\right).
\end{aligned}
$$

If $\psi(z_i) < E\psi(z_i)$ we have

$$|\psi(z_i) - E\psi(z_i)| = E\psi(z_i) - \psi(z_i) \leq (1-a^2)(1-\frac{a^2}{2}) + 2^{-\frac{3}{2}}.$$

By Hölder's inequality, we also have $s_i^2 = |X_i'\boldsymbol{\Delta}|^2 \leq \|X_i\|_\infty^2 \|\boldsymbol{\Delta}\|_1^2 \leq 16M^2 s\|\boldsymbol{\Delta}\|_2^2$, since $\boldsymbol{\Delta} \in C(S)$.
Hence
$$\frac{1}{n}|\psi(z_i) - E\psi(z_i)|s_i^2 \leq \frac{1}{n}C_a 16sM^2\|\boldsymbol{\Delta}\|_2^2, \tag{31}$$

and
$$\tilde{Z}(\nu) \leq 16 C_a s M^2 \mu^2, \tag{32}$$

with $C_a = \left[(1-a)^2(1-\frac{a^2}{2}) + 2e^{-\frac{3}{2}}\right]$.

**Step 3: Upper-bounding $E[\tilde{Z}(\nu)]$.**
Let $(\xi_i)_{i=1}^n$ be a sequence of i.i.d. Rademacher variables. It holds that

$$
\begin{aligned}
E\tilde{Z}(\nu) &= E \sup_{\boldsymbol{\Delta} \in K_\nu(S,\mu)} \left|\frac{1}{n}\sum_{i=1}^n (\psi(z_i) - E\psi(z_i))s_i^2\right| & (33) \\
&\leq 2E_{\eta,\xi} \sup_{\boldsymbol{\Delta} \in K_\nu(S,\mu)} \left|\frac{1}{n}\sum_{i=1}^n \xi_i \psi(Z_i) s_i^2\right| & (34) \\
&\leq 2(1-a^2)(1-\frac{a^2}{2}) E_{\eta,\xi} \sup_{\boldsymbol{\Delta} \in K_\nu(S,\mu)} \left|\frac{1}{n}\sum_{i=1}^n \xi_i \mathbf{1}\{z_i \leq a^2\} s_i^2\right| \\
&\quad + 4e^{-\frac{2}{3}} E_{\eta,\xi} \sup_{\boldsymbol{\Delta} \in K_\nu(S,\mu)} \left|\frac{1}{n}\sum_{i=1}^n \xi_i \mathbf{1}\{z_i > a^2\} s_i^2\right| & (35) \\
&\leq 2((1-a^2)(1-\frac{a^2}{2}) + 2e^{-\frac{2}{3}}) E_\xi \sup_{\boldsymbol{\Delta} \in K_\nu(S,\mu)} \left|\frac{1}{n}\sum_{i=1}^n \xi_i s_i(\boldsymbol{\Delta})^2\right| & (36) \\
&\leq 16M\sqrt{s}\mu((1-a^2)(1-\frac{a^2}{2}) + 2e^{-\frac{2}{3}}) E_\xi \sup_{\boldsymbol{\Delta} \in K_\nu(S,\mu)} \left|\frac{1}{n}\sum_{i=1}^n \xi_i s_i(\boldsymbol{\Delta})\right| & (37)
\end{aligned}
$$

$$
\leq 16M\sqrt{s}\mu((1-a^2)(1-\frac{a^2}{2})+2e^{-\frac{2}{3}})E_\xi \sup_{\boldsymbol{\Delta} \in K_\nu(S,\mu)} \|\boldsymbol{\Delta}\|_1 \|\frac{1}{n}\sum_{i=1}^n \xi_i X_i\|_\infty \tag{38}
$$

$$
\leq 16M\sqrt{s}\mu\nu((1-a^2)(1-\frac{a^2}{2})+2e^{-\frac{2}{3}})E_\xi \|\frac{1}{n}\sum_{i=1}^n \xi_i X_i\|_\infty. \tag{39}
$$

Here (34) follows by a standard symmetrization argument, (35) follows from simple structural results on Rademacher complexity (see e.g. [6]), (36) follows by the contraction principle (see [20]), (37) is obtained by applying Talagrand's comparison theorem (see Theorem 4.12 in [20]) noting that for any $\boldsymbol{\Delta}, \boldsymbol{\Delta}' :\in K_\nu(S,\mu)$ we have

$$
\left|\langle X_i, \boldsymbol{\Delta}\rangle^2 - \langle X_i, \boldsymbol{\Delta}'\rangle^2\right| \leq (8M\sqrt{s}\mu)\left|\langle X_i, \boldsymbol{\Delta}\rangle - \langle X_i, \boldsymbol{\Delta}'\rangle\right|,
$$

and (38) follows by Hölder's inequality.

For each $j$ the variable $\frac{1}{n}\sum_{i=1}^n \xi_i X_i^j$ is sub-Gaussian with parameter at most $M^2$. Hence we can apply exisiting bounds on the expectation of sub-Gaussian maxima (e.g., see [20]) and get

$$
E_\xi \|\frac{1}{n}\sum_{i=1}^n \xi_i X_i\|_\infty \leq 3\sqrt{2}M\sqrt{\frac{\log p}{n}}.
$$

Hence we conclude

$$
E\tilde{Z}(\nu) \leq 96 C_a M^2 \sqrt{s}\mu\nu \sqrt{\frac{\log p}{n}}, \tag{40}
$$

where $C_a = (1-a^2)(1-\frac{a^2}{2}) + 2e^{-\frac{2}{3}}$.

**Step 4: Tail bound on $\tilde{Z}(\nu)$.** In view of (31), McDiarmid's inequality implies that for any $t > 0$ we have

$$
P(\tilde{Z}(\nu) - E\tilde{Z}(\nu) \geq t) \leq \exp\left(-\frac{2nt^2}{M^4\nu^4 C_a^2}\right),
$$

where $C_a = (1-a)^2(1-\frac{a^2}{2}) + 2e^{-\frac{3}{2}}$. Let

$$
t = \frac{1}{2}\kappa_{RE}\mu^2(C_a(1-\kappa_{\lambda_\mu}) - 2e^{-\frac{3}{2}}) + C_a M^2 \mu\sqrt{s}\nu\sqrt{\frac{\log p}{n}}.
$$

Together with (40) we obtain

$$
P\Big(\tilde{Z}(\nu) \geq \frac{1}{2}\kappa_{RE}\mu^2(C_a(1-\kappa_{\lambda_\mu}) - 2e^{-\frac{3}{2}}) + 97 C_a M^2 \sqrt{s}\mu\nu\sqrt{\log(p)/n}\Big)
$$
$$
\leq \exp\Big(-\frac{2n\big(\frac{1}{2}\kappa_{RE}\mu^2(C_a(1-\kappa_{\lambda_\mu}) - 2e^{-\frac{3}{2}}) + C_a M^2\sqrt{s}\mu\nu\sqrt{\frac{\log p}{n}}\big)^2}{(16 C_a s M^2 \mu^2)^2}\Big),
$$

Hence we obtain that the event that for any $\boldsymbol{\Delta} \in K_\nu(S,\mu)$

$$
\frac{1}{n}\sum_{i=1}^n \psi(z_i) s_i^2 \leq \frac{1}{2}\kappa_{RE}\mu^2(C_a(1-\kappa_{\lambda_\mu}) - 2e^{-\frac{3}{2}}) - 97 C_a M^2 \sqrt{s}\nu\mu\sqrt{\log(p)/n} \tag{41}
$$

holds with probability at most

$$\exp\left(-n\frac{\kappa_{RE}^2(C_a(1-\kappa_{\lambda_\mu}) - 2e^{-\frac{3}{2}})^2}{2(16C_a sM^2)^2} - \frac{1}{16}\frac{\nu}{\mu s}\log p\right), \tag{42}$$

noting that for $a \geq 0, b \geq 0$, it holds that $\exp(-(a+b)^2) \leq \exp(-a^2 - b^2)$ and that $\nu/\mu > 1$, since for any $\boldsymbol{u}$, $\|\boldsymbol{u}\|_2 \leq \|\boldsymbol{u}\|_1$. By a peeling argument (see [29] for details) this yields the claim of Lemma 3. $\square$

## C Proof of Theorem 1 and Theorem 2

The proof follows the same arguments as the proof of Theorem 1 in [26]. We include details for completeness. Denote by $\delta$ the error tolerances on $\|\hat{\beta} - \beta^\star\|_2$ in the theorem statements. Define the function $F : \mathbb{R}^p \to \mathbb{R}$ as

$$F(\boldsymbol{\Delta}) = L(\boldsymbol{\beta}^\star + \boldsymbol{\Delta}) - L(\boldsymbol{\beta}^\star) + \lambda_n(\|\boldsymbol{\beta}^\star + \boldsymbol{\Delta}\|_1 - \|\boldsymbol{\beta}^\star\|_1).$$

Let $K(\delta, S) = \{\boldsymbol{\Delta} \in C(S) : \|\boldsymbol{\Delta}\|_2 = \delta\}$, where $\delta \leq \frac{\sqrt{c}}{8M\sqrt{s}}$. Let $\hat{\boldsymbol{\beta}}$ denote a minimizer of $L(\boldsymbol{\beta}) + \lambda_n\|\boldsymbol{\beta}\|_1$ in the local convexity region $\mathcal{H}_c = \{\boldsymbol{\beta}^\star + \boldsymbol{\Delta} : \|\boldsymbol{\Delta}\|_1 \leq \sqrt{c}/M\}$. Let $\hat{\boldsymbol{\Delta}} = \hat{\boldsymbol{\beta}} - \boldsymbol{\beta}^\star$. First we note that if $F(\boldsymbol{\Delta}) > 0$ for all $\boldsymbol{\Delta} \in K(\delta, S)$ then $\|\hat{\boldsymbol{\Delta}}\|_2 \leq \delta$. This is shown by contraposition. If $\|\hat{\boldsymbol{\Delta}}\|_2 > \delta$, the line joining $\hat{\boldsymbol{\Delta}}$ to $\mathbf{0}$ intersects $K(\delta, S)$ at some point $t\hat{\boldsymbol{\Delta}}$ with $t \in (0, 1)$. Since $L$ is locally convex on the line joining $(\boldsymbol{\beta}^\star + \hat{\boldsymbol{\Delta}})$ and $\boldsymbol{\beta}^\star$, by Jensen's inequality we have $F(t\hat{\boldsymbol{\Delta}}) = F(t\hat{\boldsymbol{\Delta}} + (1-t)\mathbf{0}) \leq tF(\hat{\boldsymbol{\Delta}})$. Since $F(\hat{\boldsymbol{\Delta}}) \leq 0$ then $F(t\hat{\boldsymbol{\Delta}}) \leq 0$ as well, thus showing the contrapositive statement.

To prove Theorem 1 and Theorem 2 is thus suffices to establish a lower-bound on $F(\boldsymbol{\Delta})$ over $K(\delta, S)$ for the specific values of $\delta$ in the theorem statements. For any $\boldsymbol{\Delta} \in K(\delta, S)$ due to restricted strong convexity and decomposability of the $\ell_1$-norm with respect to the set $S$, we have $F(\boldsymbol{\Delta}) \geq \langle \nabla L(\boldsymbol{\beta}^\star), \boldsymbol{\Delta} \rangle + \kappa_1 \|\boldsymbol{\Delta}\|^2 + \lambda_n(\|\boldsymbol{\Delta}_{S^c}\|_1 - \|\boldsymbol{\Delta}_S\|_1)$, where $\kappa_1$ is provided in Lemma 3. By Cauchy-Schwartz we have $|\nabla L(\boldsymbol{\beta}^\star), \boldsymbol{\Delta}\rangle| \leq \|\nabla L(\boldsymbol{\beta}^\star)\|_\infty \|\boldsymbol{\Delta}\|_1$. Notice that $\lambda_n$ in the theorems is chosen such that $\lambda_n \geq 2\|\nabla L(\boldsymbol{\beta}^\star)\|_\infty$ based on the gradient bounds of Lemmas 1 and 2. In addition $\|\boldsymbol{\Delta}\|_1 = \|\boldsymbol{\Delta}\|_S + \|\boldsymbol{\Delta}_{S^c}\|_1$. Hence we get $F(\boldsymbol{\Delta}) \geq \kappa_1 \|\boldsymbol{\Delta}\|^2 + \lambda_n(\frac{1}{2}\|\boldsymbol{\Delta}_{S^c}\|_1 - \frac{3}{2}\|\boldsymbol{\Delta}_S\|_1) \geq \kappa_1 \|\boldsymbol{\Delta}\|^2 - \frac{\lambda_n}{2}(3\|\boldsymbol{\Delta}_S\|_1)$. Since $\|\boldsymbol{\Delta}\|_1 \leq \sqrt{s}\|\boldsymbol{\Delta}\|_2$ and $\|\boldsymbol{\Delta}_S\|_2 \leq \|\boldsymbol{\Delta}\|_2$ we have $F(\boldsymbol{\Delta}) \geq \kappa_1 \|\boldsymbol{\Delta}\|^2 - \frac{\lambda_n}{2}(3\sqrt{s}\|\boldsymbol{\Delta}\|_2)$, which is strictly positive as long as $\|\boldsymbol{\Delta}\|_2 \geq \frac{1}{\kappa_1}(2\lambda_n\sqrt{s})$. The theorem statements then follow. $\square$

## D Proof of Theorem 3

Similarly as in [21], we adapt Theorem 2 in [1]. Their work assume a convex loss function. However by looking closely into their proof, one can see that the results actually rely on restricted strong convexity and restricted smoothness, and on the fact that each composite gradient update of (10) is convex. For the MD-Lasso loss we can see that the composite gradient update can be re-written as

$$\boldsymbol{\beta}^{(t+1)} = \arg\min_{\|\boldsymbol{\beta}\|_1 \leq b_0\sqrt{s}} \left(\frac{\rho}{2}\|\boldsymbol{\beta} - (\boldsymbol{\beta}^{(t)} - \frac{1}{\rho}\nabla L(\boldsymbol{\beta}^{(t)}))\|^2 + \lambda\|\boldsymbol{\beta}\|_1\right),$$

hence the results of [1] also apply to our setting, since the resticted smoothness property is satisfied by the MD-Lasso loss (in addition to the resticted convexity property stated in Lemma 3) as there exist $(\kappa_1', \kappa_2')$ such that

$$L(\boldsymbol{\beta}^\star + \boldsymbol{\Delta}) - L(\boldsymbol{\beta}^\star) - \langle \nabla L(\boldsymbol{\beta}^\star), \boldsymbol{\Delta} \rangle \leq \kappa_1' \|\boldsymbol{\Delta}\|_2^2 + \kappa_2' \sqrt{\frac{\log p}{n}} \|\boldsymbol{\Delta}\|_1^2 \qquad (43)$$

over the set $[K(S, \mu) = \{\boldsymbol{\Delta} \in C(S) : \|\boldsymbol{\Delta}\|_2 = \mu\}$. Indeed recall that

$$\nabla^2 L(\boldsymbol{\beta})(\boldsymbol{\Delta}, \boldsymbol{\Delta}) = \sum_{i=1}^n \left( \tilde{p}_i \big(1 - \frac{1}{c} r_i^2\big) s_i^2 \right) + \frac{1}{c} \Big( \sum_{i=1}^n \tilde{p}_i r_i s_i \Big)^2, \qquad (44)$$

where $\boldsymbol{\beta} = \boldsymbol{\beta}^\star + t\boldsymbol{\Delta}$, $s_i = X_i'\boldsymbol{\Delta}$, residual $r_i = (\eta_i - \langle \boldsymbol{X}_i, t\boldsymbol{\Delta} \rangle)$ and $\tilde{p}_i \propto \exp(-\frac{r_i^2}{2c})$, with $\sum_{j=1}^n \tilde{p}_j = 1$. Using similar arguments as in the proof of Lemma 3, we can show that with asymptotic probability one, for some universal positive constants $\alpha_1$ and $\alpha_2$ it holds that $\kappa_1' = \alpha_1 \kappa_{RE^u}/(1 - \kappa_{\sqrt{2c}})$ and $\kappa_2' = \alpha_2 M^2 \sqrt{s}$. Going back to Theorem 2 in [1], their so-called compound contraction coefficient specializes to

$$\delta = \Big(1 - \frac{\kappa_1}{4\kappa_1'} + \frac{64 s \kappa_2'}{\kappa_1} \sqrt{\frac{\log p}{n}}\Big)\Big(1 - \frac{64 s \kappa_2'}{\kappa_1} \sqrt{\frac{\log p}{n}}\Big)^{-1}.$$

Note that for $n$ large enough $\delta \in (0, 1)$ as $s \log p / n \to 0$. Their coefficient

$$\beta(S) = \sqrt{\frac{\log p}{n}} \Big( 2\big(\frac{\kappa_1}{4\kappa_1'} + \frac{128 s \kappa_2'}{\kappa_1} \sqrt{\frac{\log p}{n}}\big) \kappa_2 + 8\kappa_2' + 2\kappa_2 \Big),$$

which is compatible with our choice for the regularization parameter $\lambda_n$ in Theorems 1 and 2. The compound tolerance parameter as defined in [1] specializes to

$$\epsilon^2 = 8 \Big(1 - \frac{64 s \kappa'2_2}{\kappa_1} \frac{\log p}{n}\Big)^{-1} \sqrt{\frac{\log p}{n}} \Big(2\big(\frac{\kappa_1}{4\kappa_1'} + \frac{128 s \kappa_2'}{\kappa_1} \sqrt{\frac{\log p}{n}}\big) \kappa_2 + 8\kappa_2' + 2\kappa_2\Big) 6s \|\hat{\boldsymbol{\beta}} - \boldsymbol{\beta}^\star\|_2,$$

leading to the theorem statement. Similarly to the statement of Theorem 2 in [1] we have that $\big(L(\boldsymbol{\beta}^t) + \|\boldsymbol{\beta}^t\|_1\big) - \big(L(\hat{\boldsymbol{\beta}}) + \|\hat{\boldsymbol{\beta}}\|_1\big) \leq \epsilon^2/(1-\delta)$ for $t \geq \alpha_2 \frac{\log \frac{L(\boldsymbol{\beta}^{(0)}) - L(\hat{\boldsymbol{\beta}})}{\alpha_1 \|\hat{\boldsymbol{\beta}} - \boldsymbol{\beta}^\star\|_2^2}}{\log(1/\delta)}$. The theorem statement follows form the restricted strong convexity of $L$, similarly as the remark after Theorem 2 in [1]. □